\documentclass[twocolumn]{aastex631}

\usepackage{graphicx,textcomp,fancyhdr,hyperref,xcolor,fontawesome}
\definecolor{linkcolor}{rgb}{0.1216,0.4667,0.7059}
\usepackage[caption=false]{subfig}
\usepackage{float}
\usepackage{footnote}
\usepackage[noabbrev]{cleveref}

\shorttitle{Venus Zone Catalog}
\shortauthors{Colby M. Ostberg et al.}

\begin{document}

\title{The Demographics of Terrestrial Planets in the Venus Zone}

\author[0000-0002-7084-0529]{Colby Ostberg}
\affiliation{Department of Earth and Planetary Sciences, University of California, Riverside, CA 92521, USA}
\email{costb001@ucr.edu}

\author[0000-0002-7084-0529]{Stephen R. Kane}
\affiliation{Department of Earth and Planetary Sciences, University of California, Riverside, CA 92521, USA}

\author[0000-0002-4860-7667]{Zhexing Li}
\affiliation{Department of Earth and Planetary Sciences, University of California, Riverside, CA 92521, USA}

\author[0000-0002-2949-2163]{Edward W. Schwieterman}
\affiliation{Department of Earth and Planetary Sciences, University of California, Riverside, CA 92521, USA}

\author[0000-0002-0139-4756]{Michelle L. Hill}
\affiliation{Department of Earth and Planetary Sciences, University of California, Riverside, CA 92521, USA}

\author[0000-0002-4420-0560]{Kimberly Bott}
\affiliation{Department of Earth and Planetary Sciences, University of California, Riverside, CA 92521, USA}

\author[0000-0002-4297-5506]{Paul A. Dalba} 
\altaffiliation{51 Pegasi b Postdoctoral Fellow}
\affiliation{Department of Astronomy and Astrophysics, University of California, Santa Cruz, CA 95064, USA}

\author[0000-0002-3551-279X]{Tara Fetherolf}
\altaffiliation{UC Chancellor's Fellow}
\affiliation{Department of Earth and Planetary Sciences, University of California, Riverside, CA 92521, USA}

\author[0000-0003-2013-560X]{James W. Head}
\affiliation{Department of Earth, Environmental and Planetary Sciences, Brown University, Providence, RI 02912, USA}

\author[0000-0001-8991-3110]{Cayman T. Unterborn}
\affiliation{Southwest Research Institute, San Antonio, TX 78238, USA}

%%%%%%%%%%%%%%%%%%%%%%%%%%%%%%%%%%%%%%%%%%%%%%%%%%%%%%%%%%%%%%%%%%%%

\begin{abstract}

Understanding the physical characteristics of Venus, including its atmosphere, interior, and its evolutionary pathway with respect to Earth, remains a vital component for terrestrial planet evolution models and the emergence and/or decline of planetary habitability. A statistical strategy for evaluating the evolutionary pathways of terrestrial planets lies in the atmospheric characterization of exoplanets, where the sample size provides sufficient means for determining required runaway greenhouse conditions. Observations of potential exoVenuses can help confirm hypotheses about Venus' past, as well as the occurrence rate of Venus-like planets in other systems. Additionally, the data from future Venus missions, such as DAVINCI, EnVision, and VERITAS, will provide valuable information regarding Venus, and the study of exoVenuses will be complimentary to these missions. To facilitate studies of exoVenus candidates, we provide a catalog of all confirmed terrestrial planets in the Venus Zone, including transiting and non-transiting cases, and quantify their potential for follow-up observations. We examine the demographics of the exoVenus population with relation to stellar and planetary properties, such as the planetary radius gap. We highlight specific high-priority exoVenus targets for follow-up observations including: TOI-2285~b, LTT~1445~A~c, TOI-1266~c, LHS~1140~c, and L~98-59~d. We also discuss follow-up observations that may yield further insight into the Venus/Earth divergence in atmospheric properties.

\end{abstract}


%%%%%%%%%%%%%%%%%%%%%%%%%%%%%%%%%%%%%%%%%%%%%%%%%%%%%%%%%%%%%%%%%%%%

\section{Introduction}
\label{sec:intro}

Exoplanets have been discovered at an extraordinary rate over the past several decades, with thousands of confirmed exoplanets now known \citep{akeson2013}. This large sample of known planets has motivated studies of exoplanet populations/demographics \citep{tremaine2012,ford2014,winn2015} and comparison to the Solar System architecture \citep{limbach2015,martin2015b,horner2020b,kane2021d}. Although initial discoveries were dominated by giant planets \citep{butler2006,udry2007}, subsequent surveys and improved instrumentation have enabled the detection of terrestrial planets. Radial velocity (RV) surveys have improved dramatically in understanding noise sources and increasing precision \citep[e.g.,][]{pepe2014a,fischer2016,stefansson2016,gupta2021}, and legacy RV data have allowed the detection of long-period planets \citep[e.g.,][]{fischer2014a,butler2017,trifonov2020,fulton2021,rosenthal2021}. In parallel, space-based transit surveys, such as the Kepler mission \citep{borucki2010a} and the Transiting Exoplanet Survey Satellite (TESS) \citep{ricker2015}, have pushed the sensitivity of exoplanet experiments deeply into the terrestrial regime.

A significant challenge in characterizing the terrestrial planet population is the modeling of their potential surface conditions \citep{way2017b,fauchez2021,wolf2022}. A subset of the discovered terrestrial exoplanets will have system properties that makes them amenable to follow-up observations to study their atmospheres \citep{morley2017b,batalha2018b,lincowski2019,lustigyaeger2019a}. These atmospheric studies are crucial for ascertaining the true nature of the planetary surface properties and its evolutionary pathway. A direct comparison is generally made to our local analog of Venus, whose differing atmospheric evolution from Earth has been the subject of considerable amounts of research effort \citep{donahue1982,kasting1988c,hamano2013,way2016,kane2019d,way2020,turbet2021}. This research can be furthered by studying Earth-like and Venus-like exoplanets to identify whether incident flux is primarily responsible for the climate divergence of Venus, or if it is due to other factors, such as planet size, degassing rates, and atmospheric loss.

Atmospheric spectroscopy is likely to produce the most robust diagnostic that can distinguish between the various exoVenus scenarios \citep{ehrenreich2012a,barstow2016telling}, and various methods have been developed to predict the potential signal-to-noise of transmission spectroscopy observations \citep{kempton2018,ostberg2019}. With a vast array of exoplanet targets now available, and coupled with limited follow-up opportunities, the prioritization of terrestrial targets is becoming increasingly important. Categorizing the known planets as lying in the Habitable Zone (HZ) of their host star \citep{kasting1993a,kopparapu2013a,kopparapu2014}, the region around a star where surface water may be present on terrestrial planets given sufficient atmospheric pressure, is an effective means to creating such a prioritization scheme, with a focus on potential Earth analogs. To aid in these efforts, various catalogs of HZ planets have been constructed \citep{hill2023catalog}, including those based on Hipparcos data \citep{chandler2016}, Kepler discoveries \citep{kane2016c}, and the TESS observational strategy \citep{kaltenegger2019b}. Given the intrinsic biases of exoplanet detection methods toward close in planets, many more exoplanets lie within the Venus Zone (VZ) of their system, defined as the region around a star interior to the runaway greenhouse boundary, and thus where terrestrial planets may be Venus analogs in a post runaway greenhouse state \citep{kane2014e}.

In the near-term, the primary method for studying the atmospheric composition of Venus-like worlds will be through transmission spectroscopy, which is used to determine the wavelengths at which light is absorbed when passing through a planet’s atmosphere. Venus’ transmission spectrum was modeled in preparation for Venus’ stellar occultation in 2012, which demonstrated that the Venusian cloud and haze layers prevent transmission spectroscopy from probing the atmosphere below an altitude of 80~km \citep{ehrenreich2012a}. The effect of clouds on Venus' transmission spectrum was also shown to cause difficulties for retrieval algorithms that were unable to consistently differentiate Earth and Venus-like atmospheres \citep{barstow2016telling}. \citet{Lincowski2018} modeled the transmission spectra of the TRAPPIST-1 planets assuming they had 10-bar Venus-like atmospheres. Their work illustrated that the weaker CO$_2$ absorption bands at 1.05 and 1.3 $\mu$m and absorption caused by sulfuric acid clouds are likely to be the best avenues for determining if a planet has Venus-like surface conditions. Simulated JWST observations of the TRAPPIST-1 planets with Venus-like atmospheres showed that their atmospheres could be detected in less than 20 transit observations, but discerning their compositions would take more than 60 transit observations \citep{lustigyaeger2019a}. The catalog and selection criteria provided in this work will help to guide the identification of VZ planets for atmospheric follow-up observations to test atmospheric evolution scenarios.

Here, we present the results of a compilation and analysis of VZ candidates from the known inventory of exoplanets. The purpose of the VZ catalog is to study the demographics of terrestrial planets that lie within the VZ, and to facilitate the prioritization of follow-up targets whose atmospheric characterization will provide critical diagnostics in determining the inner edge of the HZ. Section~\ref{sec:data} describes the extraction and parsing of exoplanet data, the radius constraints used to identify terrestrial planets, and the calculations that allow the analyses of the bulk data characteristics. The creation of an extensive table of terrestrial planets whose orbits lie partially or wholly within the VZ is presented in Section~\ref{sec:catalog}. In Section~\ref{sec:discussion}, we provide a discussion of the VZ population demographics and highlight potential priority for James Webb Space Telescope (JWST) targets, as well as other important follow-up opportunities. Since the catalog includes both transiting and non-transiting exoplanets, we further discuss the importance of non-transiting targets for future studies of terrestrial planet evolution. Section~\ref{sec:conclusions} includes a summary of the main results, concluding remarks, and an outline of further potential uses of the VZ catalog.

%%%%%%%%%%%%%%%%%%%%%%%%%%%%%%%%%%%%%%%%%%%%%%%%%%%%%%%%%%%%%%%%%%%%

\section{Data Extraction and Calculations}
\label{sec:data}
 
%Suggestions: First subsection on data extraction from NEA, including parsing of data, filters (such as eccentricity) applied, etc. Second subsection on HZ calculations, application of M-R relationships, etc.

%Radius cutoff reasoning from Lincowski et al (2018): There is growing evidence that the bulk properties of planets with radii . 1:5R  are more similar to our Solar System terrestrials than to sub-Neptune planets with H2-dominated atmospheres (e.g. Rogers 2015; deWit et al. 2016; Fulton et al. 2017; deWit et al. 2018).

We define terrestrial VZ planets to be planets with radii $R_p < 2.0 $~$R_\oplus$ that spend any portion of their orbits within the boundaries of the VZ. Note that the radius limit for terrestrial planets depends on numerous factors, including formation scenarios and composition \citep{unterborn2019}, and not all planets within this range will indeed be rocky \citep{rogers2015}. However, this radius cutoff was chosen to account for uncertainties in radius measurements or calculations to minimize the exclusion of any terrestrial planets from the sample. Terrestrial VZ planets with measured radii have an average uncertainty of 0.2 R$_\oplus$, while the maximum uncertainty is as high as 1.03 R$_\oplus$. Planets with radii $R_p > 2.0 $~$R_\oplus$ and relatively high or low mass likely require significant Fe-enrichment relative to their host star or surface volatiles to explain their anomalously higher or lower than expected density. These planets would therefore be more likely to be super-Mercuries or volatile-rich mini-Neptunes/water worlds, respectively \citep{unterborn2016,unterborn2019}. 

\begin{figure*}
  \includegraphics[width = 0.95\textwidth]{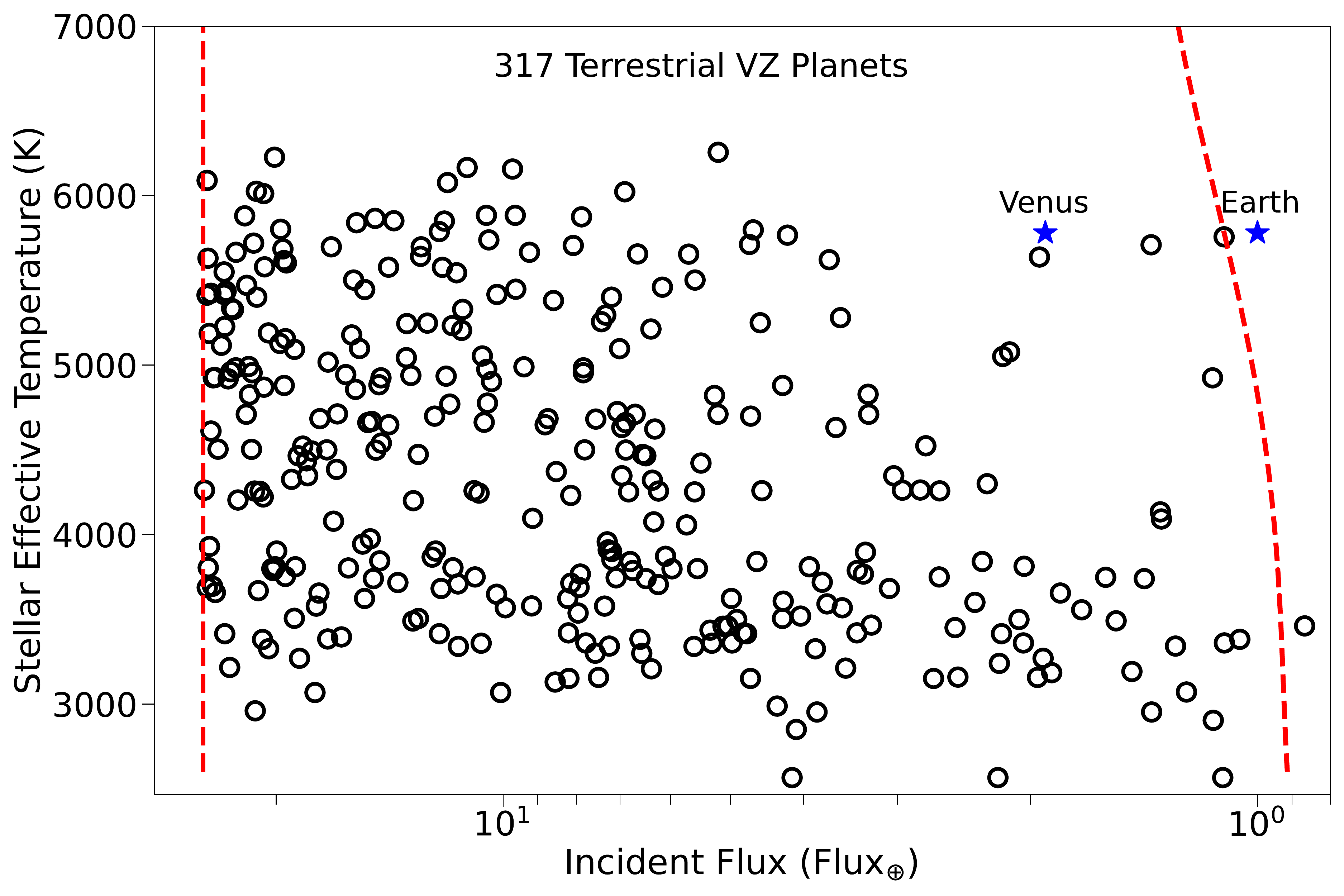}
  \caption{Terrestrial planets from the NEA that spend any part of their orbit in the VZ. The inner VZ boundary is indicated by the left line, and the right line is the outer VZ boundary. Earth and Venus are shown for reference (blue stars).}
  \label{fig:VZ_Planets}
\end{figure*}

The inner boundary of the VZ is defined as $25\times$ the incident flux received by Earth ($F_\oplus$), which is the insolation flux limit where Venus would begin to lose the majority of its atmosphere \citep{zahnle2017}. Note that this does not account for variations in stellar activity as a function of spectral type, but is broadly encapsulating the vast range of expected stellar ages and masses, as well as planetary masses and atmospheres, expected to lie within our sample. The outer VZ boundary is the runaway greenhouse boundary, which corresponds to the inner boundary of the conservative HZ (CHZ), and is defined as the insolation flux threshold where liquid water on Earth's surface would be evaporated, forcing it into a runaway greenhouse state \citep{kopparapu2013a,kopparapu2014}. To determine whether a planet orbits within the VZ we used Kepler's equation to calculate the planet's distance from its host star as a function of its orbit. If a planet's distance is ever less than the distance of the outer VZ boundary and greater than the distance of the inner VZ boundary, then the planet orbits within the VZ. The 317 known terrestrial planets from the NEA that spend any amount of their orbit in the VZ (hereafter referred to as VZ terrestrial planets) are shown in Figure~\ref{fig:VZ_Planets}. It should be noted that planets within the VZ are not guaranteed to have Venus-like surface conditions. The VZ is instead a first-order estimate for identifying planets that may be Venus-like and for guiding target selection for follow-up observations with JWST or other future facilities. These observations will be the primary method of producing more accurate predictions of the surface conditions of VZ planets.

The possibility of habitable worlds within the VZ cannot be discounted either, as it has been shown that Venus could have maintained temperate surface conditions for as recently as 1~Gya \citep{way2020}. However, this scenario requires a young Venus to be cool enough to allow water to condense on its surface, which faces challenges due to the possible lack of cloud formation at the substellar point \citep{turbet2021day}. Atmospheric spectroscopy of terrestrial planets in the VZ will be essential for investigating both the possibility of a temperate period in Venus' history, and the conditions that force a planet into a runaway greenhouse. Furthermore, terrestrial planets have long been known to differ significantly in their geologic \citep{head1977geologic} and tectonic evolution \citep{head1981tectonic}, and their mechanisms of global lithospheric heat transfer and loss \citep{head1981topography} suggested the presence of ongoing global plate tectonics \citep{head1987evidence}, a hypothesis tested by the Magellan mission. Magellan data revealed a geologically very young surface \citep[< ~1 Gyr;][]{mckinnon1997cratering} and a single global lithospheric plate lacking a system of ongoing  plate tectonics \citep{solomon1991venus,solomon1992venus}. Similarly, the currently observed atmosphere of Venus may have formed relatively recently geologically, transitioning from Earth-like oceanic conditions \citep{way2020} (the ‘great climate transition’) during the early stages \citep{byrne2021venus} of the young remnant surface geological record \citep{ivanov2011a}, or in an additive manner throughout this period \citep{khawja2020tesserae}.  Alternatively, the current Venus atmosphere may be a “fossil atmosphere” dating from some time earlier in the history of Venus \citep{head2021contributions}. Indeed, the conundrum of the Earth-Venus relationship \citep{head2014geologic} is one of the most compelling outstanding scientific questions today. As we explore both forward models and inverse models to understand the nature, origin and evolution of the Venus atmosphere, the greatly enlarged exploratory parameter space provided by the demographics of  terrestrial exoplanets in the Venus Zone is designed to contribute a much needed broader perspective on the problem, and a guide to obtaining the most critical observations to improve our understanding.

All of the planetary and stellar data used in this work were acquired from the NASA Exoplanet Archive \citep[NEA;][]{akeson2013} using the Application Program Interface (API). We used the default properties for each system, and the data are current as of 2023 January 19 \citep{nea}. Each planetary system was required to have non-null values for the host star effective temperature and the planetary semi-major axis, or the means to calculate it, else it was removed from the sample. If orbital eccentricity was not measured for a planet, the orbit was assumed to be circular, while the argument of periastron was set to 90\degr \ if no value was available. The majority of VZ planets have an orbital eccentricity of 0, with 76\% of them having no measured eccentricity which were then assumed to be 0, while 6\% of planets were observed to have 0 eccentricity. The remaining planets all had non-zero eccentricities. Assuming circular orbits may have excluded planets which would enter the VZ with eccentric orbits, however we expect this to have a negligible effect on the total amount of terrestrial VZ planets. This assumption planets beyond the outer VZ boundary, however a circular orbit is the appropriate assumption for planets between the inner VZ boundary and the host star because of the likelihood of tidal locking.

If not available for a given system, the values for stellar luminosity, planet incident flux, and planet equilibrium temperature were calculated when possible. If either a measured planet radius or mass were not available, then the missing values were calculated using the methodology of \citet{chen2017}. Due to the restraints of mass-radius relationship, radius calculations were limited to planets with mass $M_p < 25$~$M_\oplus$, while mass calculations required planet radius $R_p < 5$~$R_\oplus$. We did not incorporate the uncertainties when calculating the mass or radius using the \citet{chen2017} method. The estimated RV amplitude for each planet was calculated when possible, using equations 12--14 from \citet{lovis2010radial}. If the semi-major axis of the planet was not available, it was calculated from the orbital period and stellar mass.

%%%%%%%%%%%%%%%%%%%%%%%%%%%%%%%%%%%%%%%%%%%%%%%%%%%%%%%%%%%%%%%%%%%%

\section{A Catalog of VZ Exoplanets}
\label{sec:catalog}

Here we present the data for all known VZ planets that were identified using the methodology described in Section~\ref{sec:data}. These results are shown in Table~\ref{tab:planets}, where the data are listed in alphabetical order based on the planet names. The percentage of a planet's orbit that is spent within the VZ and CHZ are shown in the columns labeled `VZ (\%)' and `CHZ (\%)', respectively. If a planet does not spend 100\% of its orbit in the VZ, then it will spend the remainder of its orbit in either the CHZ, or between the inner VZ boundary and the host star. Thus, the summation of the CHZ and VZ columns will generally constitute 100\% of the orbit, except in cases where the planet ventures interior to the inner VZ boundary.

The location column, marked `Loc', indicates where the relative location of the planet's semi-major axis, $a$, lies within the VZ. The location values range from 0--1, where 0 is the inner VZ boundary and 1 is the outer VZ boundary. Planetary radii ($R_p$) and masses ($M_p$) are included in the table, with italicized values representing those that were calculated using the \citet{chen2017} mass-radius relationship. Orbital period, $P$, semi-major axis, $a$, planetary equilibrium temperature, $T_\mathrm{eq}$, and RV amplitude, $K$, are included for each planet when available. We included the calculated transmission spectroscopy metric (TSM) value for each planet using the methods of \citet{kempton2018}. The TSM is used to estimate the achievable signal-to-noise ratio (S/N) of transmission spectroscopy with JWST assuming 10 hours of observations. In this work we do not focus on the magnitude of individual TSM values, but instead on TSM values in reference to each other. In general, we view the planets with the largest TSM values compared to other planets in the sample as having the most potential for success in follow-up observations. The orbital eccentricity, $e$, of a planet was set to 0 when no eccentricity measurement was available, which are listed as a single italicized 0 in the table. Measured eccentricities of 0 are listed as `0.000'.

Table~\ref{tab:stars} includes details for each VZ planet's host star. $R_\star$ and $M_\star$ are the radius and mass of the star in solar units, respectively. Jmag and Vmag are the J-band and V-band magnitude of the star, respectively. The stellar effective temperature, $T_\mathrm{eff}$, surface gravity, $\log g$, luminosity, $L_\star$, and distance, $d$, of the stars are also shown in Table~\ref{tab:stars}. In the case of the NEA providing null values for $L_\star$, we calculated $L_\star$ using $R_\star$ and $T_\mathrm{eff}$ when they are available.

%%%%%%%%%%%%%%%%%%%%%%%%%%%%%%%%%%%%%%%%%%%%%%%%%%%%%%%%%%%%%%%%%%%%

\section{Discussion}
\label{sec:discussion}

%%%%%%%%%%%%%%%%%%%%%%%%%%%%%%%%%%%%%%%%%%%%%%%%%%%%%%%%%%%%%%%%%%%%

\subsection{Demographics}
\label{sec:demo}

\begin{figure*}
\centering 
\subfloat{%
  \includegraphics[width=0.95\columnwidth, height=0.8\columnwidth]{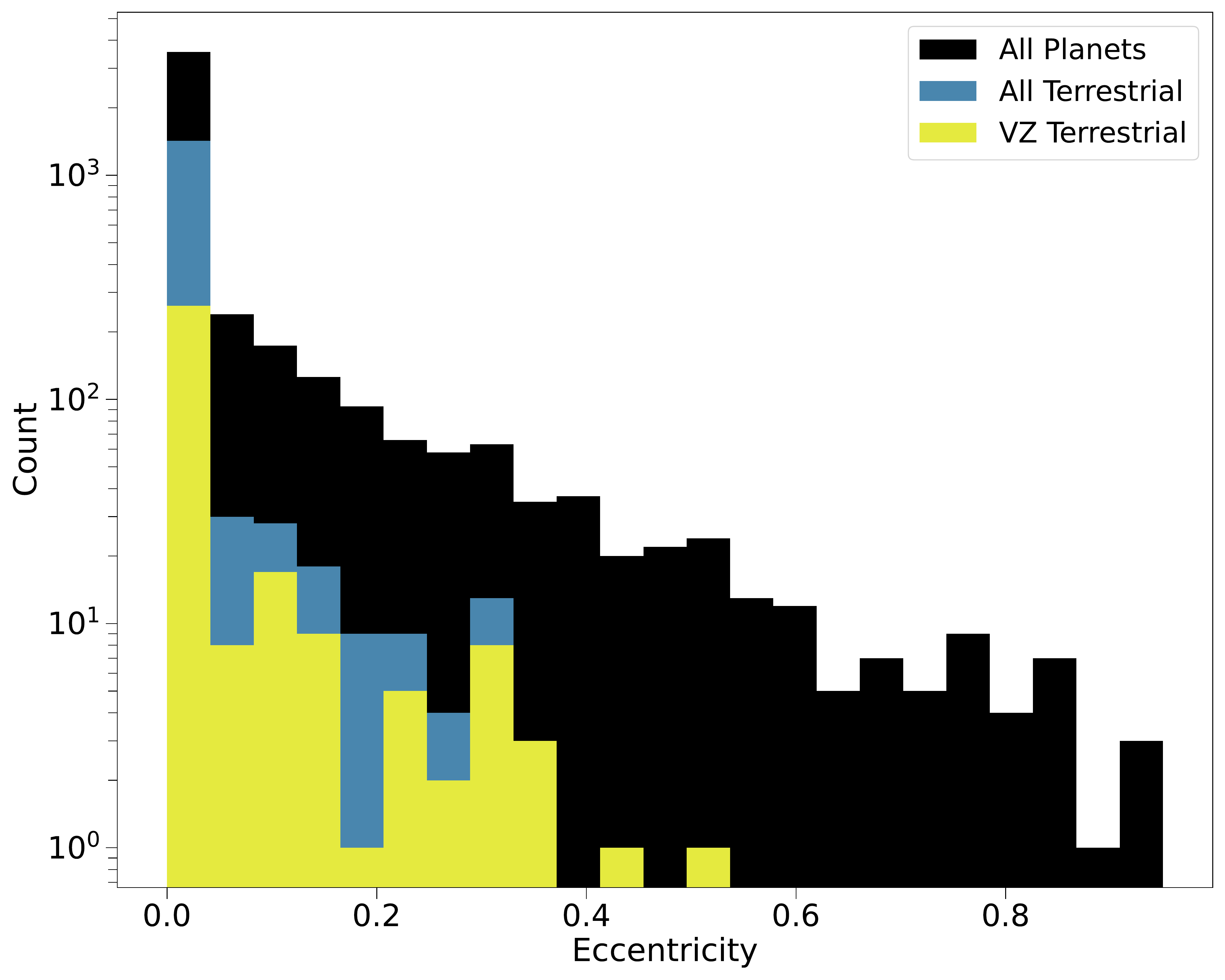}%
}
\subfloat{%
  \includegraphics[width=0.95\columnwidth, height=0.8\columnwidth]{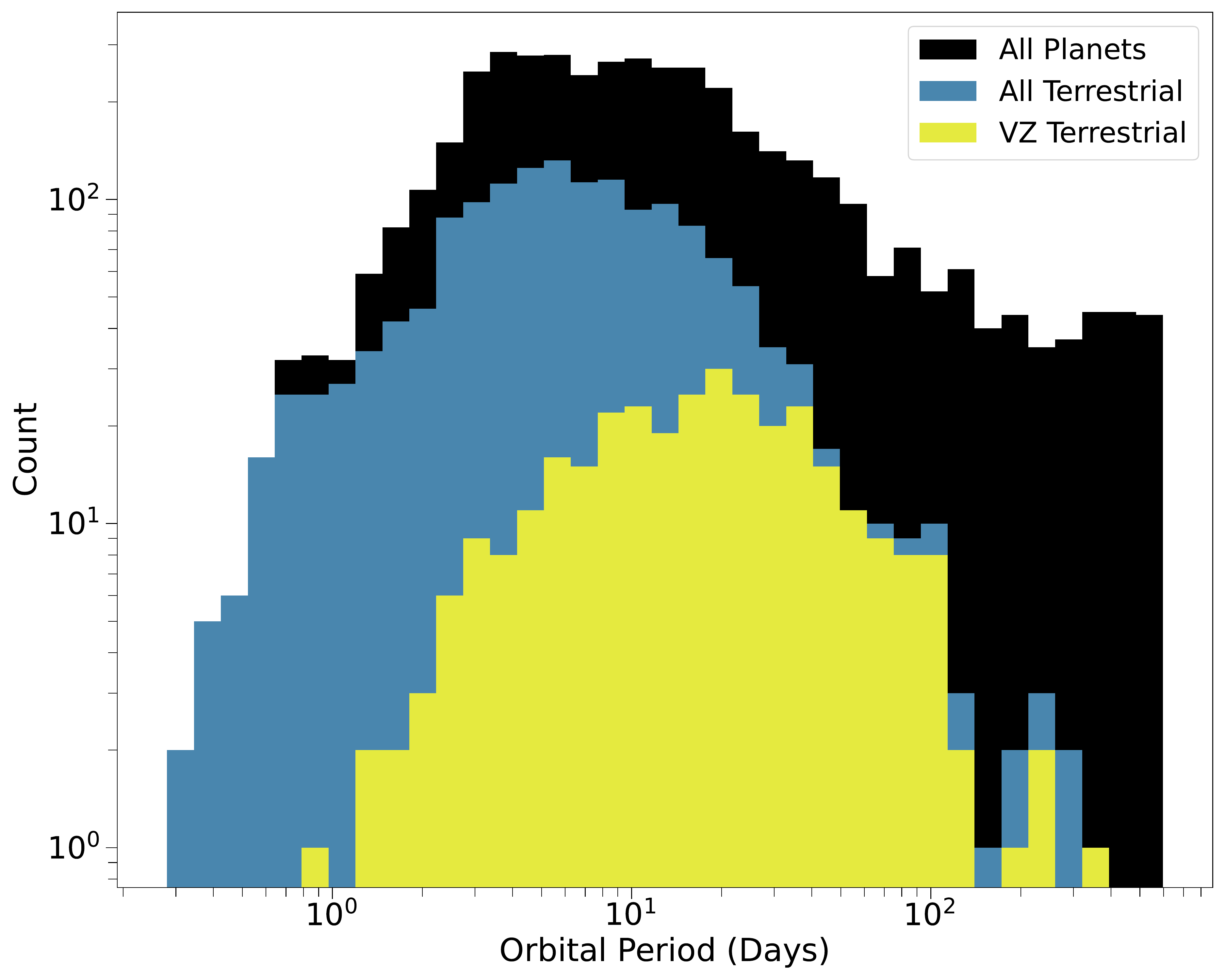}%
}\qquad
\subfloat{%
  \includegraphics[width=0.95\columnwidth, height=0.8\columnwidth]{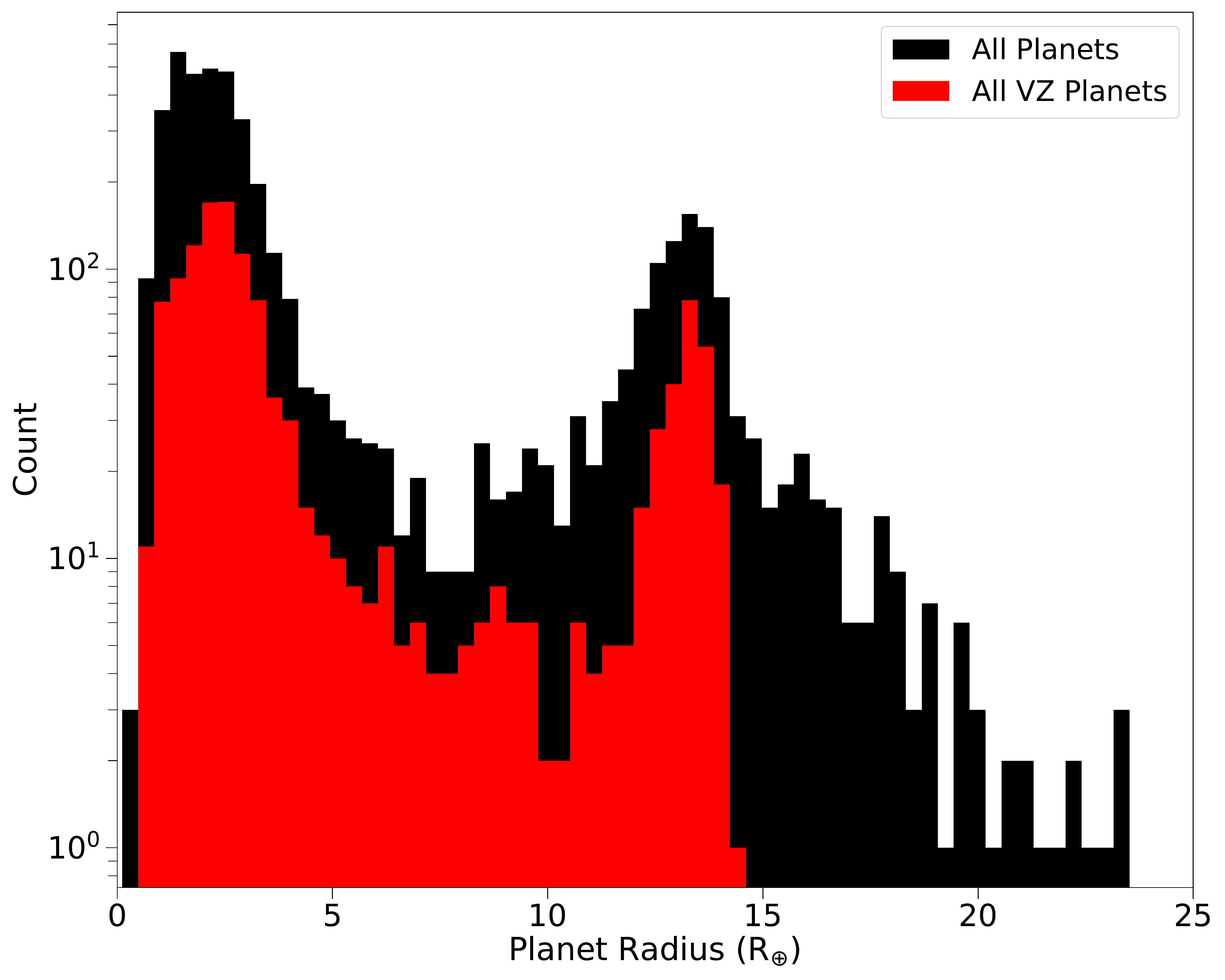}%
}
\subfloat{%
  \includegraphics[width=0.95\columnwidth, height=0.8\columnwidth]{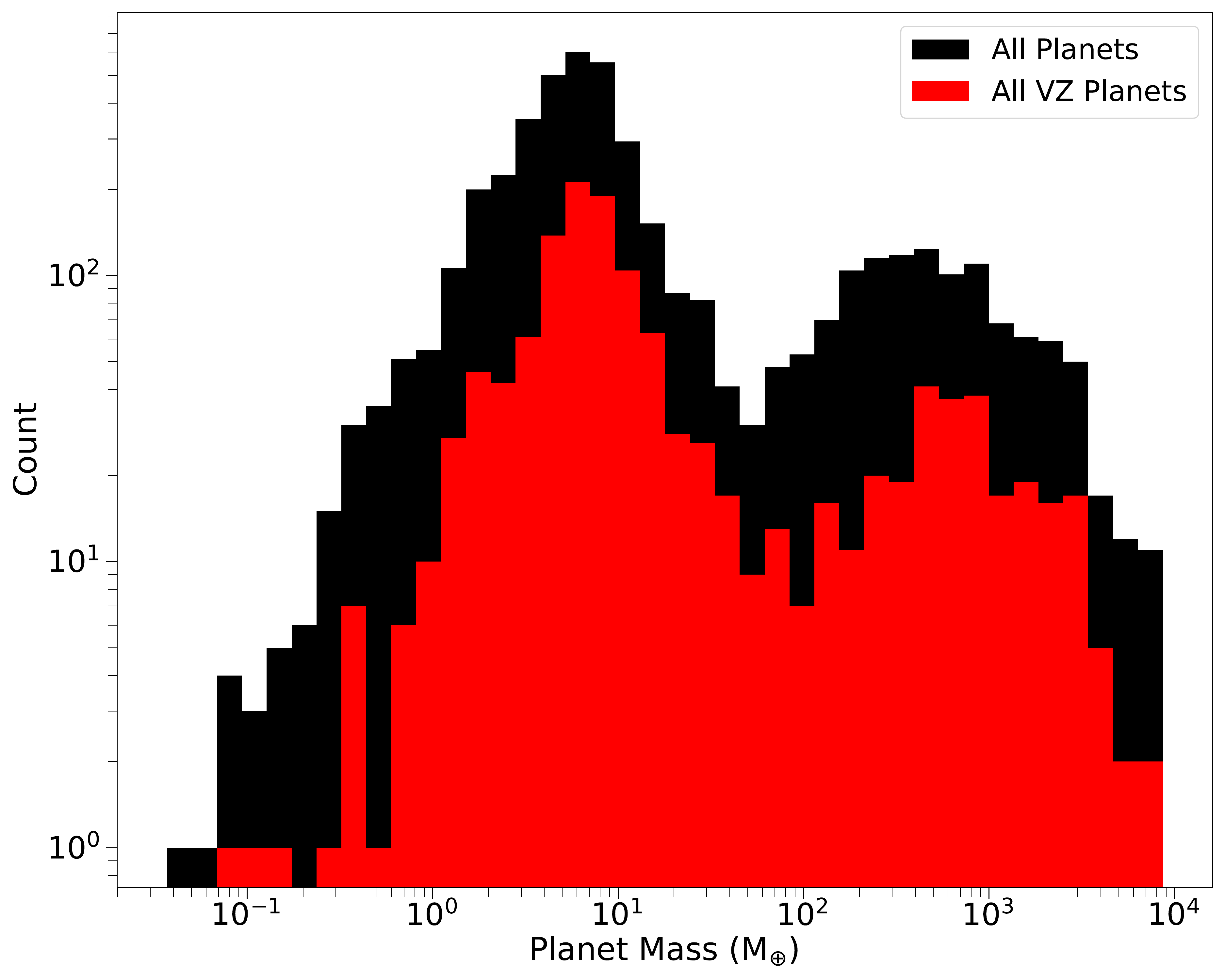}%
}
\caption{Histograms illustrating the distribution of orbital eccentricity (upper left), orbital period (upper right), planet radius (bottom left), and planet mass (bottom right). In the upper two plots, the yellow distribution indicates terrestrial VZ planets and the blue distribution is all known terrestrial planets. In the bottom two plots the red distribution is all known planets within the VZ, regardless of their size. In all four plots the black distribution is all known planets regardless of radius or location.
\label{fig:planet_hist}}
\end{figure*}

Unveiling the demographics of various planet populations provides insight into planet formation and evolution scenarios, which are particularly important for the potential creation of runaway greenhouse planets. Figure~\ref{fig:planet_hist} presents property distributions for the planets in our sample, including orbital eccentricity (upper left), orbital period (upper right), planet radius (lower left), and planet mass (lower right). The shown distributions compare the VZ terrestrial planets (yellow) with all terrestrial planets (blue) and the general planet population (black). For the VZ terrestrial (yellow) and all terrestrial (blue) populations, the period distribution is remarkably gaussian (at least in log space), such that the distribution peaks approximately at the mean value. However, this is not true of the general planet population, which is rather asymmetric due to inclusion of cold giant planets. Interesting to note that the orbital period limit of the terrestrial planet distribution approximately coincides with the outer edge of the VZ. This is due to planet detection missions being biased towards closer in planets which has resulted in the discovery of only 29 terrestrial planets with orbits beyond the outer VZ boundary \citep{hill2023catalog}. The planet radius and mass histograms compare VZ planets of all radii to all known exoplanets. A radius gap is visible in both the VZ and all planet distributions \citep[e.g.][]{lopez2013role,owen2013kepler,fulton2017}, with a sharp cutoff of VZ planets at $\sim$14~$R_\oplus$. The cutoff is present because the gravitational force of high mass planets typically prevents their size from exceeding $\approx$ 14 R$_\oplus$. Exceptional cases such as hot--Jupiters can exceed this limit because their atmospheres puff up from high insolation flux, however these planets are too close to their stars to be in the VZ.

\begin{figure*}
\centering 
\subfloat{%
  \includegraphics[width=0.95\columnwidth]{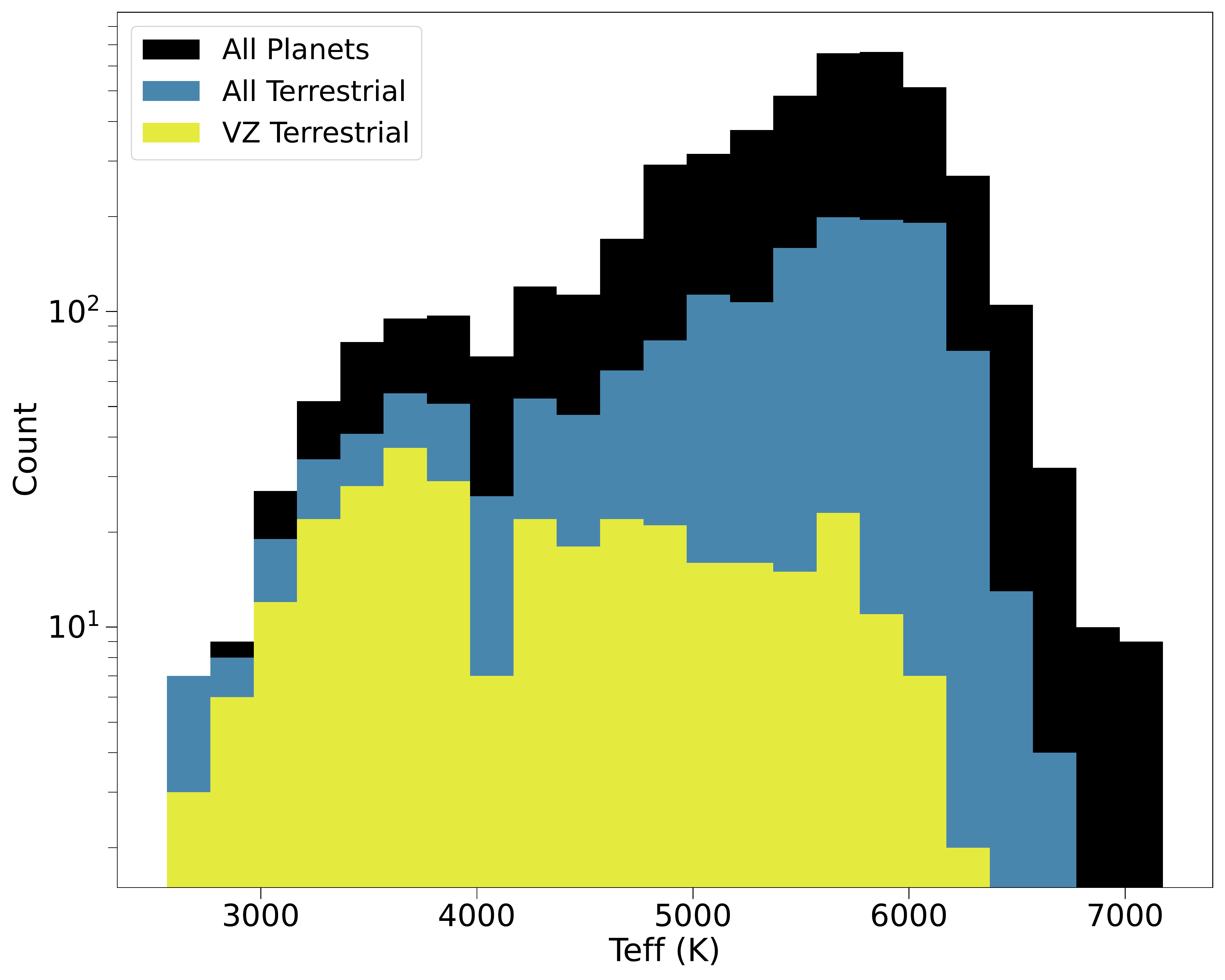}%
}
\subfloat{%
  \includegraphics[width=0.95\columnwidth]{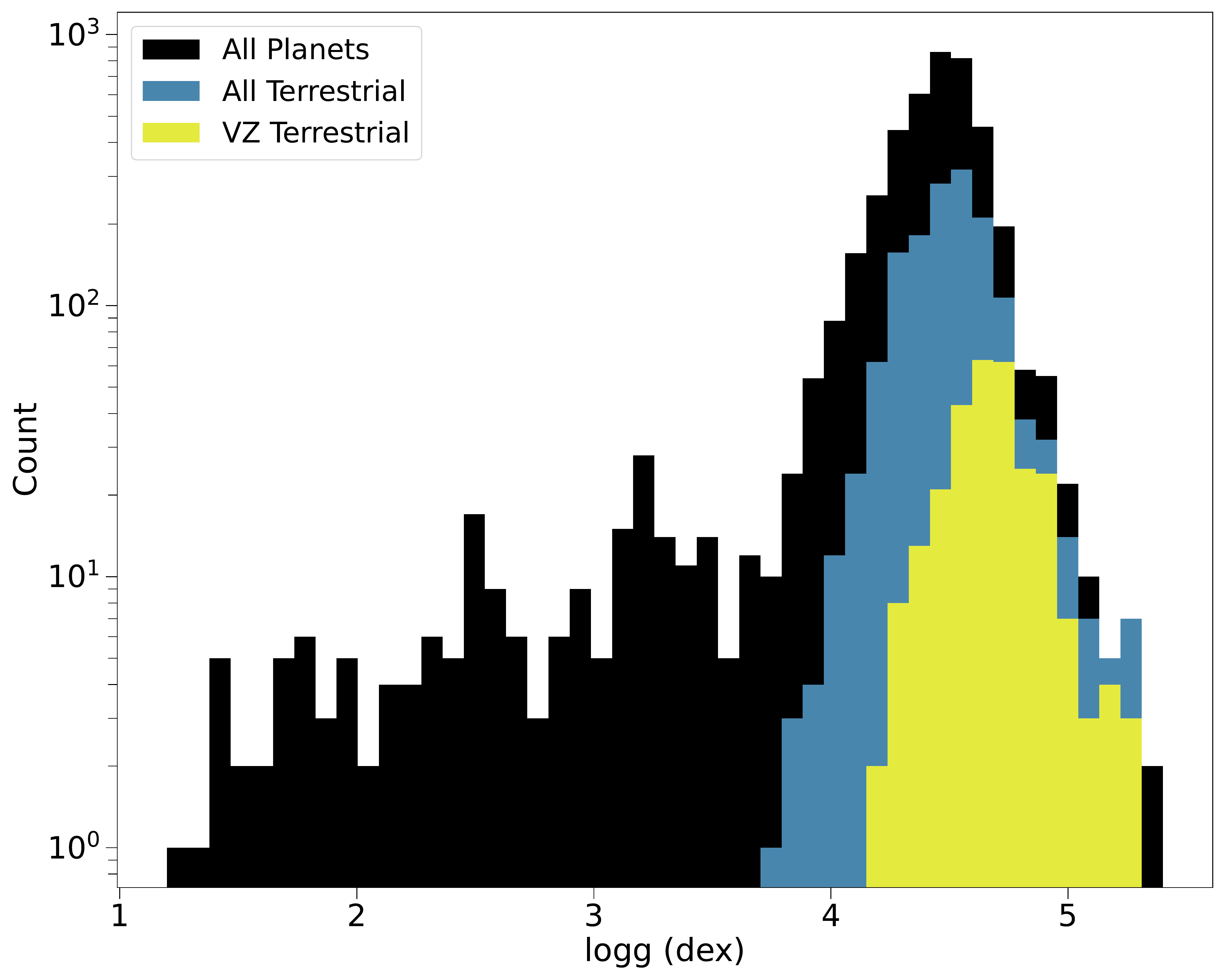}%
}\qquad
\subfloat{%
  \includegraphics[width=0.95\columnwidth]{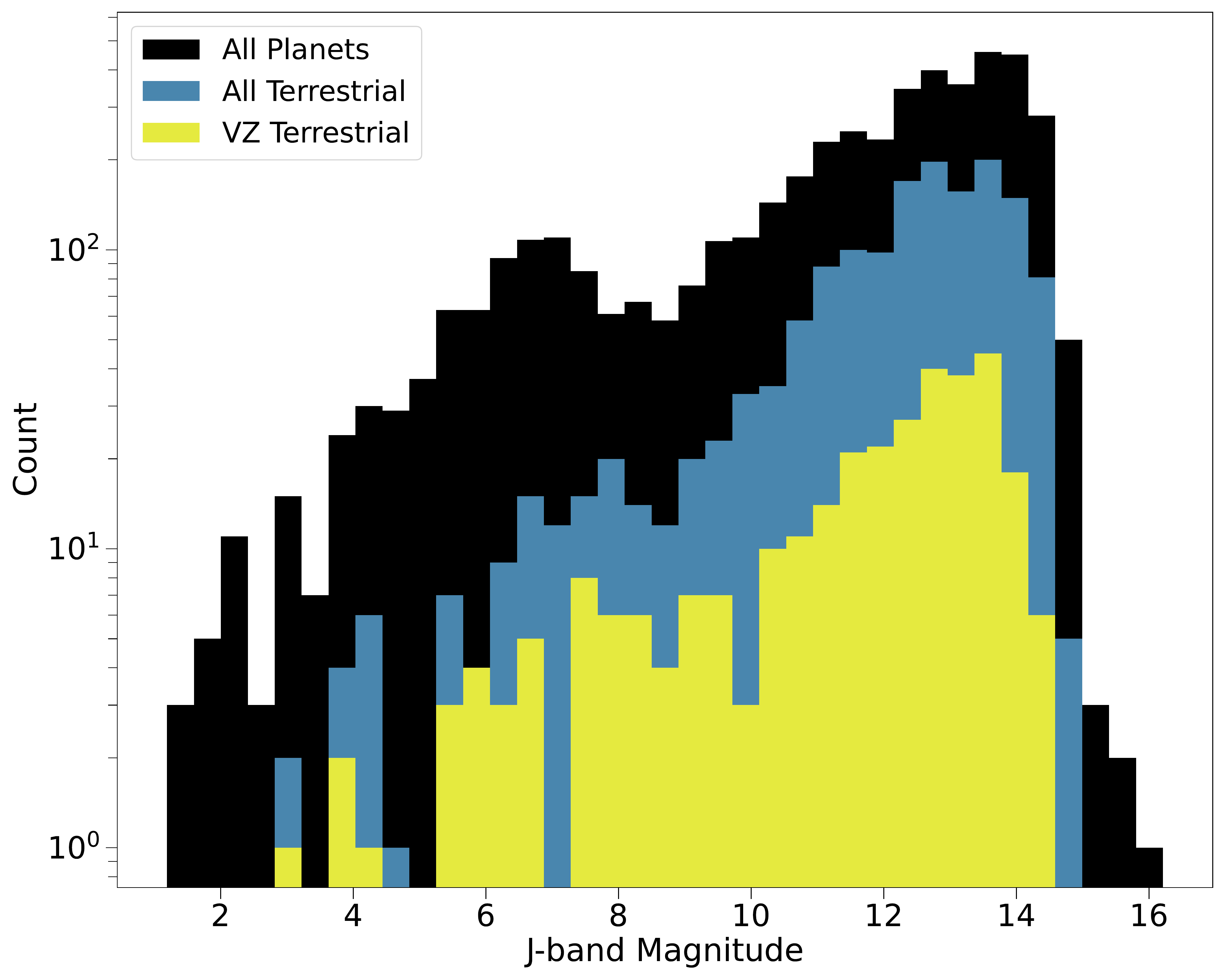}%
}
\subfloat{%
  \includegraphics[width=0.95\columnwidth]{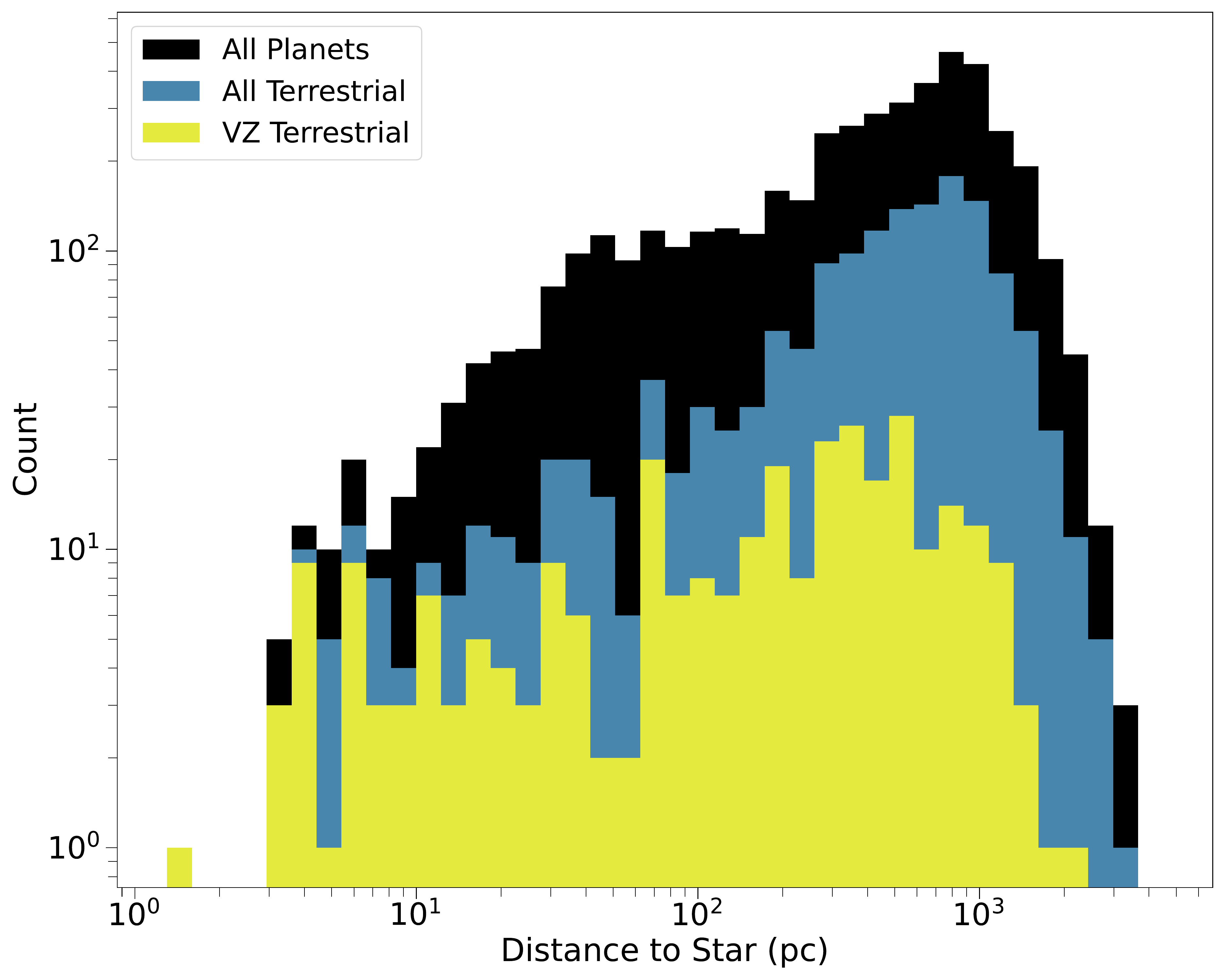}%
}
\caption{Histograms displaying distributions of stellar properties: stellar effective temperature (upper left), surface gravity (upper right), J-band magnitude (bottom left), distance to star (bottom right). In yellow is the distribution for the host stars of terrestrial VZ planets, in blue is the host stars of all terrestrial planets, and the black distribution is all known planets regardless of size or location.
\label{fig:star_hist}}
\end{figure*}

Host star properties such as $T_\mathrm{eff}$ (upper left),  $\log g$ (upper right), Jmag (lower left), and stellar distance (lower right) are displayed in Figure~\ref{fig:star_hist}. All plots compare the host stars of VZ terrestrial planets to those of all terrestrial planets, and all planets regardless of their size. Terrestrial VZ planets (yellow) are most commonly found around cooler, smaller stars where the sensitivity of observations toward terrestrial planets is greater, and is thus likely related to an observational bias. By contrast, the two other subsets (black and blue) have been more commonly detected around hotter, brighter Sun-like stars that are better suited for RV observations. Similarly, the dependence of the respective detected planet populations on $\log g$ demonstrates that the terrestrial planets, including those in the VZ, are overwhelmingly found around main sequence stars, whose relatively small size enable the detection of proportionally small planets. Furthermore, the majority of exoplanets have been found around fainter stars, with the distribution of host star J-band magnitudes peaking at $\sim$13. Note that the J-band magnitude distribution for the sample that includes all exoplanet host stars has an additional peak at $\sim$7, largely due to the effect of RV surveys that target preferentially brighter stars than those included in transit surveys. The bottom-right panel of Figure~\ref{fig:star_hist} shows the distribution of stellar distances for the respective samples, where the number of known planets naturally rises with increasing distance due to the volume limited nature of the various exoplanet surveys. Note also the indication of the nearest exoplanet, Proxima Centatui b, at the far left of the figure; a non-transiting exoVenus candidate (see Section~\ref{sec:nontransit}).

\begin{figure*}
\centering 
\subfloat{%
  \includegraphics[width=0.95\columnwidth]{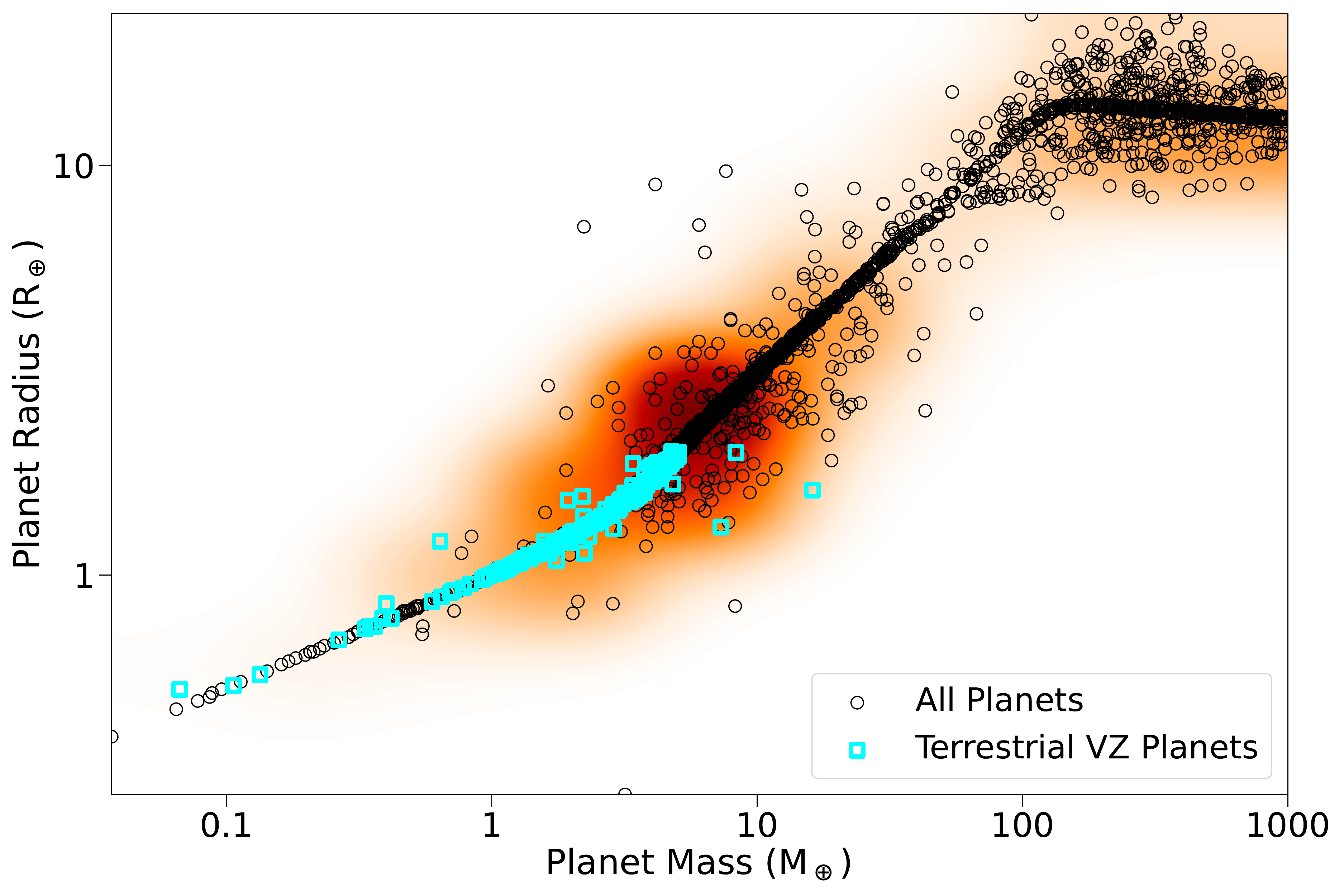}%
}
\subfloat{%
  \includegraphics[width=0.95\columnwidth]{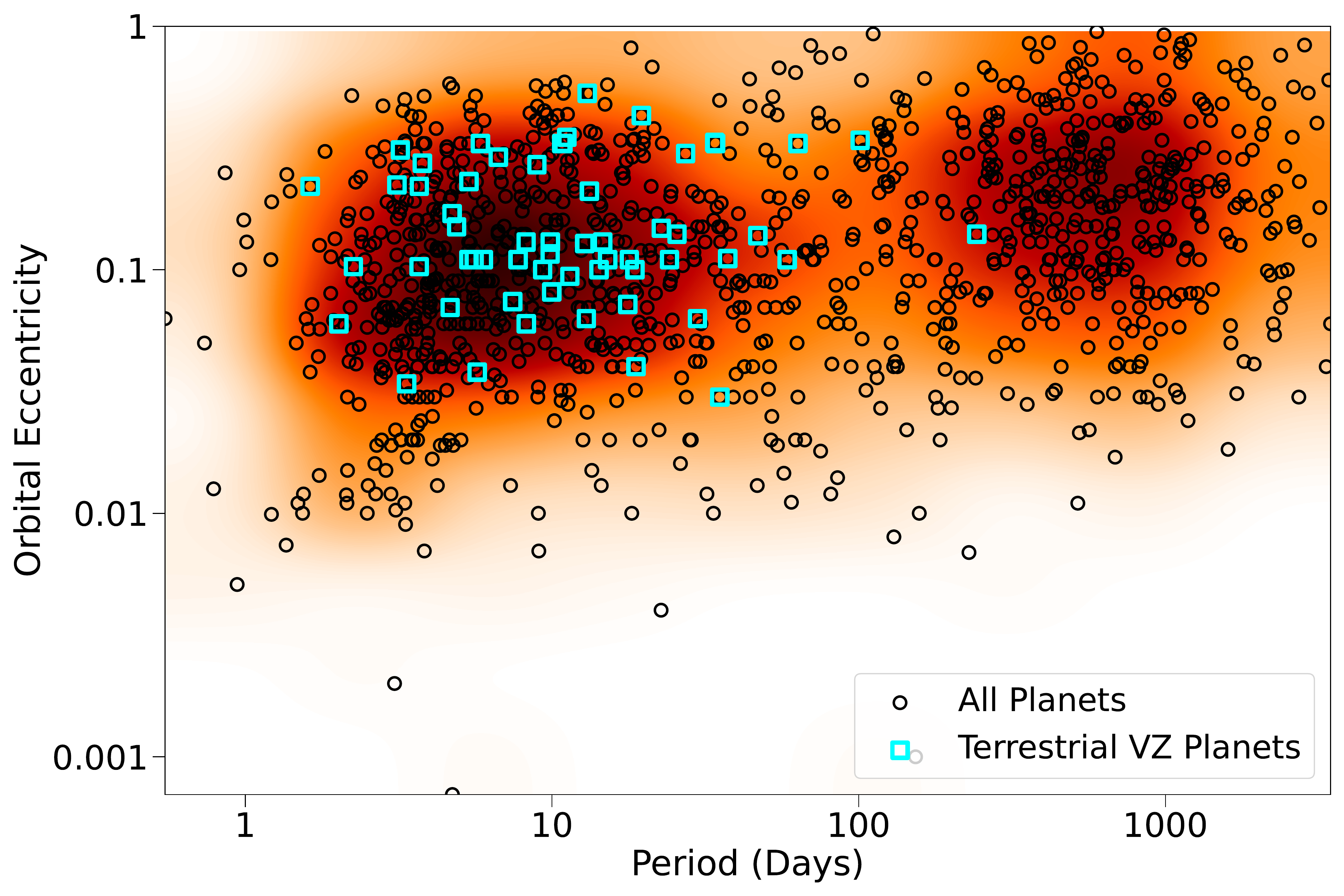}%
}\qquad
\subfloat{%
  \includegraphics[width=0.95\columnwidth]{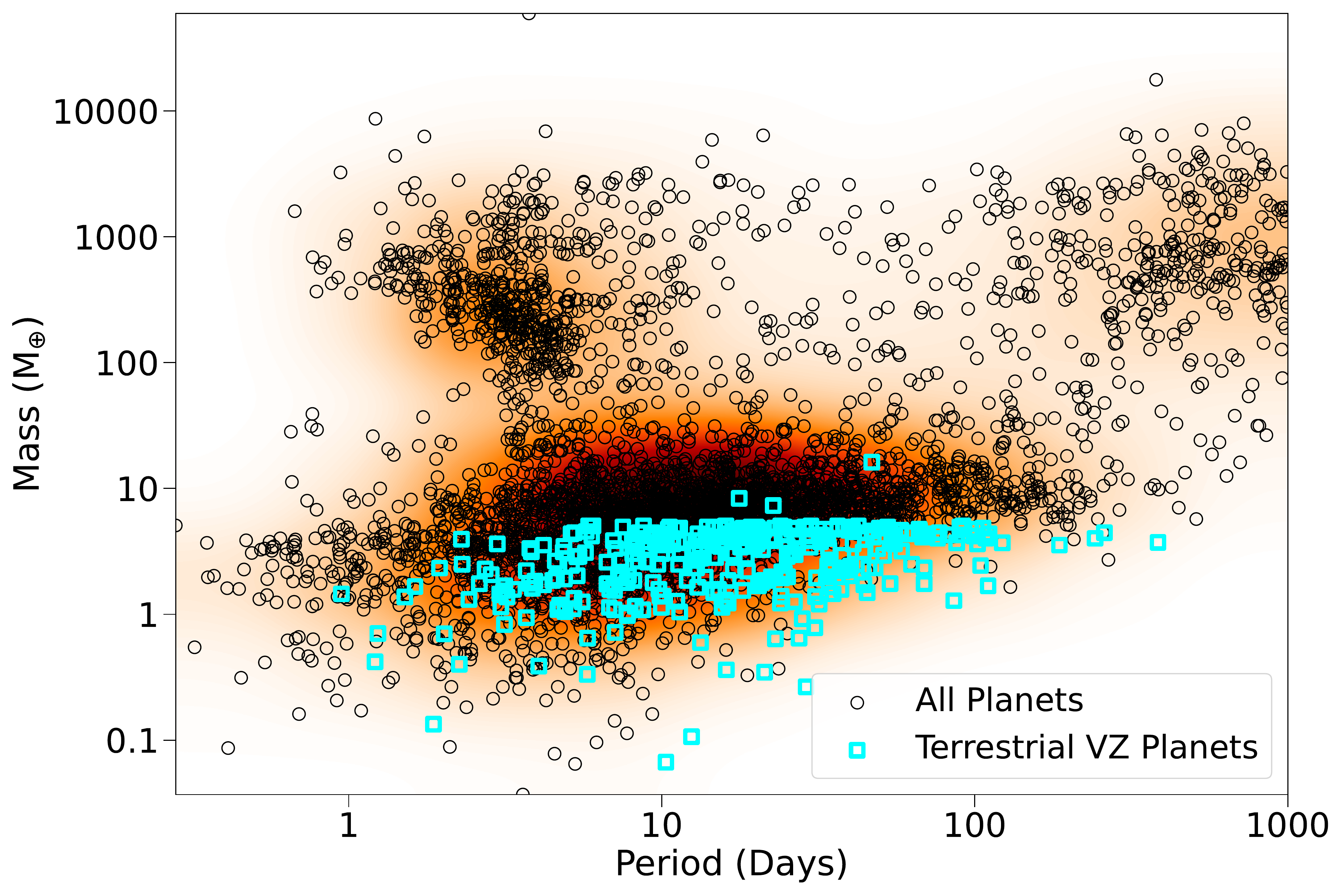}%
}
\subfloat{%
  \includegraphics[width=0.95\columnwidth]{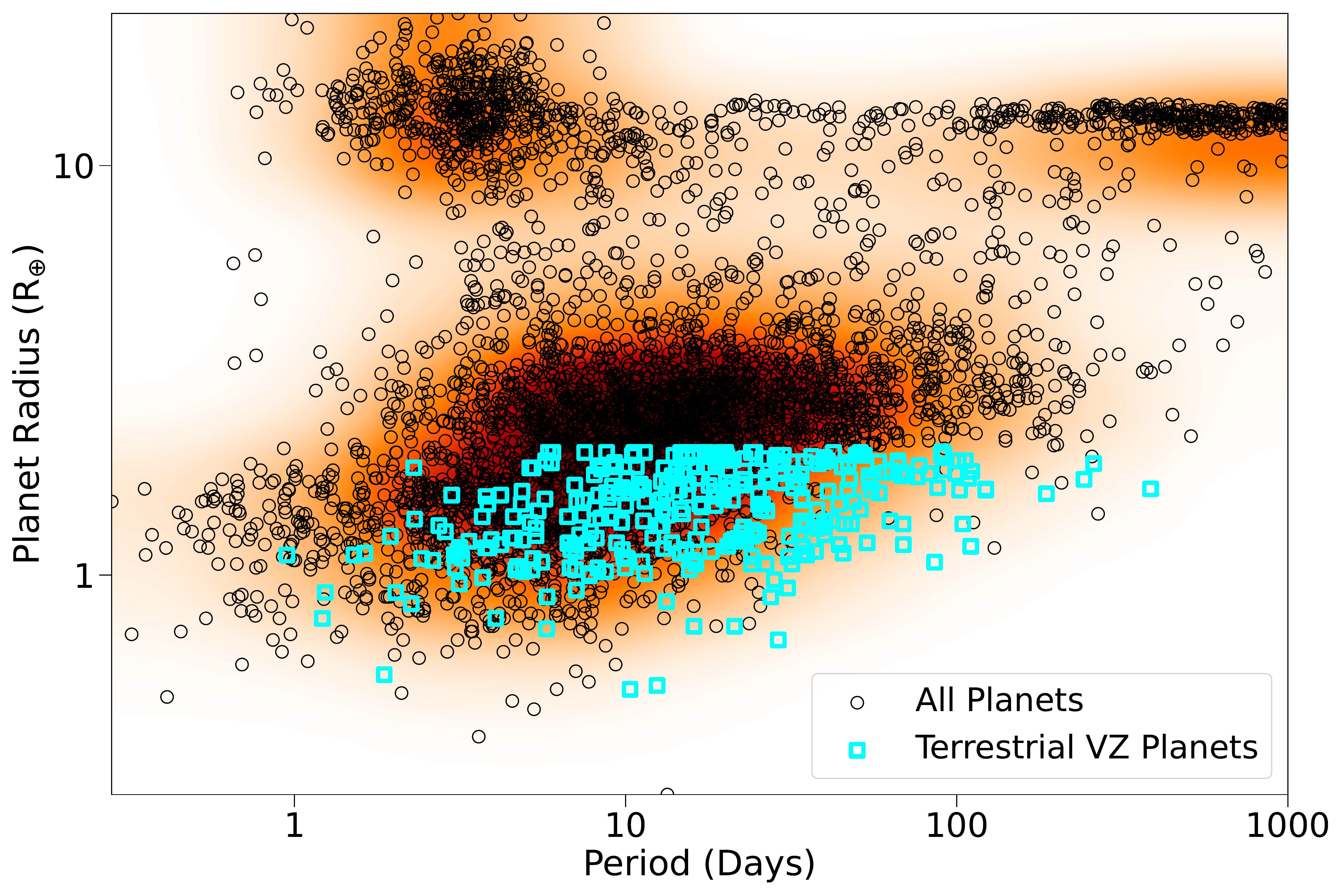}%
}
\caption{The relationships between various planet properties. In all plots, the black circles are all known planets regardless of size or location while the blue squares are terrestrial VZ planets.
\label{fig:scatter_plots}}
\end{figure*}

Figure~\ref{fig:scatter_plots} displays relationships between various parameters for terrestrial VZ planets compared to all known planets. The upper left figure shows planet mass versus planet radius. Since most VZ planets lack either a radius and mass measurement, the trend seen is representative of the mass-radius relationship that was used to calculate the missing mass or radius values \citep{chen2017}. The upper right figure displays orbital period vs orbital eccentricity for all planets with non-zero eccentricities. Few VZ planets are displayed since the majority did not have measured eccentricities. The lower left plot displays orbital period vs planet mass and the lower right shows orbital period vs planet radius. The upper limit planet radius set for our VZ sample can be seen in the lower right plot. Similarly, there is a cutoff in mass in the lower left plot caused by how most masses were calculated using the mass-radius relationship. There is an additional radius cutoff for all known planets around 10 $R_\oplus$ which is also caused by the mass-radius relationship. The relationship plateaus at higher planet mass and radius as demonstrated in the upper left plot.

\setlength{\belowcaptionskip}{2pt}

\setlength{\belowcaptionskip}{2pt}

\setlength{\belowcaptionskip}{2pt}

%%%%%%%%%%%%%%%%%%%%%%%%%%%%%%%%%%%%%%%%%%%%%%%%%%%%%%%%%%%%%%%%%%%%

\subsection{JWST VZ Targets}
\label{sec:jwst}

Observations of exoplanet atmospheres will provide information that is necessary for understanding their climates. During Cycle~1, JWST will conduct such observations for eight terrestrial VZ planets: GJ~357~b \citep{luque2019planetary}, GJ~1132~b \citep{bonfils2018radial}, TRAPPIST-1~b \citep{agol2021refining}, TRAPPIST-1~c \citep{agol2021refining}, TOI-776~b \citep{luque2021planetary}, TOI-776~c \citep{luque2021planetary}, LTT~1145~A~b \citep{winters2022second}, and L~98-59~b \citep{demangeon2021warm}. TOI-776~c has a radius of 2.02~$R_\oplus$, but is within the upper radius limit for terrestrial planets when considering uncertainties. All of these planets will be observed with either the Mid-Infrared Instrument \citep[MIRI;][]{wright2004jwst}, Near-Infrared Spectrograph \citep[NIRSpec;][]{bagnasco2007overview}, or the Near-Infrared Slitless Spectrograph \citep[NIRISS;][]{doyon2012jwst} aboard JWST. MIRI will be primarily used to observe the emission spectra of exoplanets, while NIRISS and NIRSpec will conduct transmission spectroscopy. 

TOI-776~b, TOI-776~c, and L~98-59~b have the 3 highest TSM values among the VZ JWST targets. However, they are in close vicinity to the inner VZ boundary, which may have led to significant atmospheric loss \citep{van2018asteroseismic}. TOI-776~b and TRAPPIST-1~c VZ planets are intriguing as they both have additional terrestrial VZ planets within their respective systems. TOI-776~b is accompanied by TOI-776~c, and TRAPPIST-1~c, b and d are all in the VZ. Atmospheric spectroscopy of multiple VZ planets within the same system offers the opportunity to investigate how VZ planet evolution may vary based on location in the VZ. Furthermore, the TRAPPIST-1 VZ planets will also be useful for comparison to their neighboring planets in the conservative HZ, TRAPPIST-1~e, f and g. Transmission or emission spectroscopy of VZ and HZ exoplanets in the same system would be a unique opportunity to investigate evolutionary differences between Earth and Venus using exoplanets. TRAPPIST-1~c will receive 6 transit and 4 eclipse observations, while TRAPPIST-1~b, TOI-776~b, and TOI-776~c will each receive 2 transit observations. 

GJ~357~b has a radius of 1.22~$R_\oplus$ and is located 0.01 AU (Loc = 0.093) from the inner VZ boundary. The close proximity to its host star causes GJ~357~b to experience and incident flux of $\sim$13~$F_\oplus$, placing it at risk of sustaining significant atmospheric loss. Transmission spectroscopy of the planet will aid in determining the presence of an atmosphere and, in turn, be a useful test for the location of the inner VZ boundary. GJ~357~b will be observed for a single transit by JWST~NIRSpec during Cycle~1 observations. GJ~1132~b is also located close to the inner VZ boundary, and its eccentric orbit results in it spending 23\% of the orbit between the inner VZ boundary and its host star. It has a radius of 1.13~$R_\oplus$, making it also susceptible to substantial atmospheric loss and a useful test for the location of the inner VZ boundary as well. GJ~1132~b is anticipated to have two transit observed by JWST NIRSpec during Cycle~1.

LTT~1145~A~b orbits a single star but is in an M-dwarf triple binary system with LTT 1145 BC. The system the planet is in makes it unique from the other VZ planets being observed by JWST, however since LTT 1145 BC are 34 AU from LTT 1145 A, it is unlikely that they have any significant contribution to the energy budget of LTT~1145~A~b. There are two sets of observations planned for LTT~1145~A~b, with a single transit observation by JWST NIRSpec and three secondary eclipse observations by JWST MIRI. Additional observations of its neighboring terrestrial planet in the VZ would illustrate whether their differences in received flux resulted in different climates. 

L~98-59 b spends 63\% of its orbit in the VZ, and the rest between the inner VZ boundary and its host star. Throughout its eccentric orbit the planet receives stellar flux that varies from 19--29~$F_\oplus$. The likelihood that the planet has sustained significant atmospheric loss is increased by both its high incident flux and smaller size of 0.85~$R_\oplus$. Two transit observations with JWST NIRSpec are planned for the planet, and detection of an atmosphere on this planet would demonstrate that VZ planets can sustain an atmosphere closer to the star than predicted by the inner VZ boundary. Besides L~98-59~b, there are 3 other terrestrial planets in its system that are also in the VZ. Observations of all planets in the system would be valuable for understanding the runaway greenhouse effect across a range of planet scenarios for the same host star. 

The success of these observations will be highly dependent on the planets' atmospheres. If they were to have Venus-like atmospheres, then it is likely that JWST will be unable to detect any molecular species in their transmission spectra given the allotted observation time, but may still be able to determine the presence of an atmosphere. If the planets have Earth-like or oxygen-desiccated atmospheres, then detecting molecules in their atmospheres may be acheived in as little as 6 transits \citep{lustigyaeger2019a, pidhorodetska202198}.

%%%%%%%%%%%%%%%%%%%%%%%%%%%%%%%%%%%%%%%%%%%%%%%%%%%%%%%%%%%%%%%%%%%%

\subsection{VZ Targets for Future Observations}
\label{sec:followup}

\begin{figure}
  \includegraphics[width = 0.95\columnwidth]{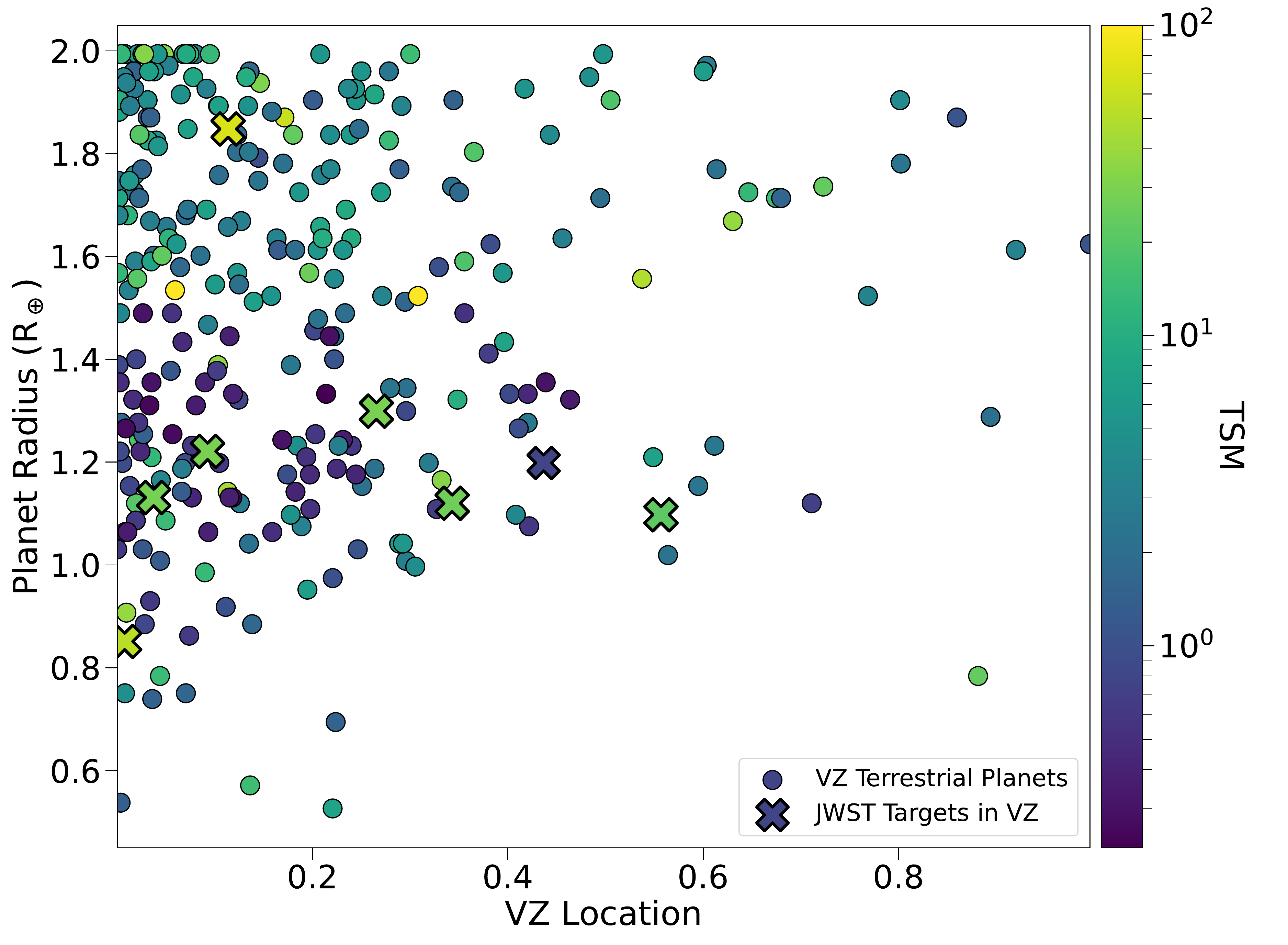}
  \caption{Terrestrial VZ planets colored based on their TSM value. The x-axis represents the planets' location in the VZ, where 1 indicates the outer VZ boundary and 0 is the inner VZ boundary. Data points denoted by an 'x' are VZ planets that are planned to be observed by JWST.}
  \label{fig:TSM}
\end{figure}

Here we highlight the VZ planets with high scientific value which are the most amenable to atmospheric spectroscopy and should be considered for observations in JWST Cycle 2+ or by other future facilities. All planets we considered were required to be known to transit, and the TSM value was used as the basis for quantifying the observational potential of planets. A high equilibrium temperature can cause a planet to have a high TSM value, but planets closer to the outer VZ boundary maximize similarities between the planets and Venus. Figure~\ref{fig:TSM} displays the VZ location and radii of VZ planets, with points colored based on their TSM value. Based on our selection criteria, an optimal planet would be at 0.95~$R_\oplus$, at the far right of the figure, and colored yellow. There is a scarcity of planets in that region, and therefore no planets that exactly match our criteria. Hence, we handpicked the planets we deemed to have the best combination of observation potential, size similarity to Venus, and general scientific promise. The five planets we determined to have the best combination of scientific intrigue and observational amenability are TOI-2285~b \citep{fukui2022toi}, LTT~1445~A~c \citep{winters2022second}, TOI-1266~c \citep{demory2020super}, LHS~1140~c \citep{lillo2020planetary}, and L~98-59~d \citep{demangeon2021warm}.  

TOI-2285~b is the only planet of the six that spends a portion (12\%) of its orbit in the CHZ due to its orbital eccentricity of 0.3. The TSM of this planet is the second lowest of the group, but its orbit makes it an interesting test case for the climate of a planet in both the VZ and HZ. Out of the 6 planets, LTT~1445~A~c is the closest in size and mass to that of Venus, with a measured radius and mass of 1.14~$R_\oplus$ and 1.54~$M_\oplus$, respectively. The planet orbits a single star in an M-dwarf triple binary with LTT~1445~BC and has a TSM of 45, which is the third highest of the group. 
TOI-1266~c is in the center of the VZ, and has the second highest TSM value of the six targets. It has a measured radius of 1.55~$R_\oplus$ and a measured mass of 2.2~$M_\oplus$, making it considerably larger and more massive than Venus. Observing its atmosphere would help determine whether the surface conditions of more massive planets in the VZ differ from those closer to Venus in size. In addition, the detection or non-detection of an atmosphere on TOI-1266~c would help justify the location of the inner VZ boundary by illustrating whether planets more massive than Venus can sustain an atmosphere despite having a higher equilibrium temperature and orbiting a highly active M-dwarf star. 

LHS~1140~c is the second most similar to Venus in size from the six future targets, with a measured radius of 1.16~$R_\oplus$. The planet orbits an ultra-cool red-dwarf, but its orbital period of 3.7 days results in an incident flux of 4.3~$F_\oplus$ and an equilibrium temperature of 709~K, which is the hottest of the group. Determining whether the planet has an atmosphere would help inform the atmospheric escape experienced by Earth-sized VZ planets with high incident flux, which will test the location of the inner VZ boundary. 

L~98-59~d is in a heavily populated VZ with three other terrestrial planets. Planet d was chosen from this system as it is the farthest planet from the inner VZ boundary that is also known to transit. L~98-59~d has by far the largest TSM value in the group and the third highest TSM value of all VZ planets. The large amount of VZ planets and their amenability to atmospheric spectroscopy make the L 98-59 system an important benchmark for studying the VZ.

The high priority targets of particular interest are those that have a neighboring terrestrial planet or planets in the VZ or HZ. LTT~1445~A~c has a single terrestrial neighbor in the VZ that is planned to be observed by JWST, while L 98-59 d has three neighboring VZ planets, one of which will be observed by JWST. Currently the VZ only provides a general estimate for the climates of terrestrial planets, but observations of multiple VZ planets within the same system can be used to study the evolutionary differences of planets in the inner and outer VZ of the same system. LHS~1140~c is the only planet in the group with a neighboring terrestrial planet in the HZ. Observations of both LHS~1140~c and b provide a chance to compare the differences between Earth and Venus to planets in a similar system.

%%%%%%%%%%%%%%%%%%%%%%%%%%%%%%%%%%%%%%%%%%%%%%%%%%%%%%%%%%%%%%%%%%%%

\subsection{Non-Transiting Planets of Interest}
\label{sec:nontransit}

\begin{figure*}
\centering 
\subfloat{%
  \includegraphics[width=0.95\columnwidth]{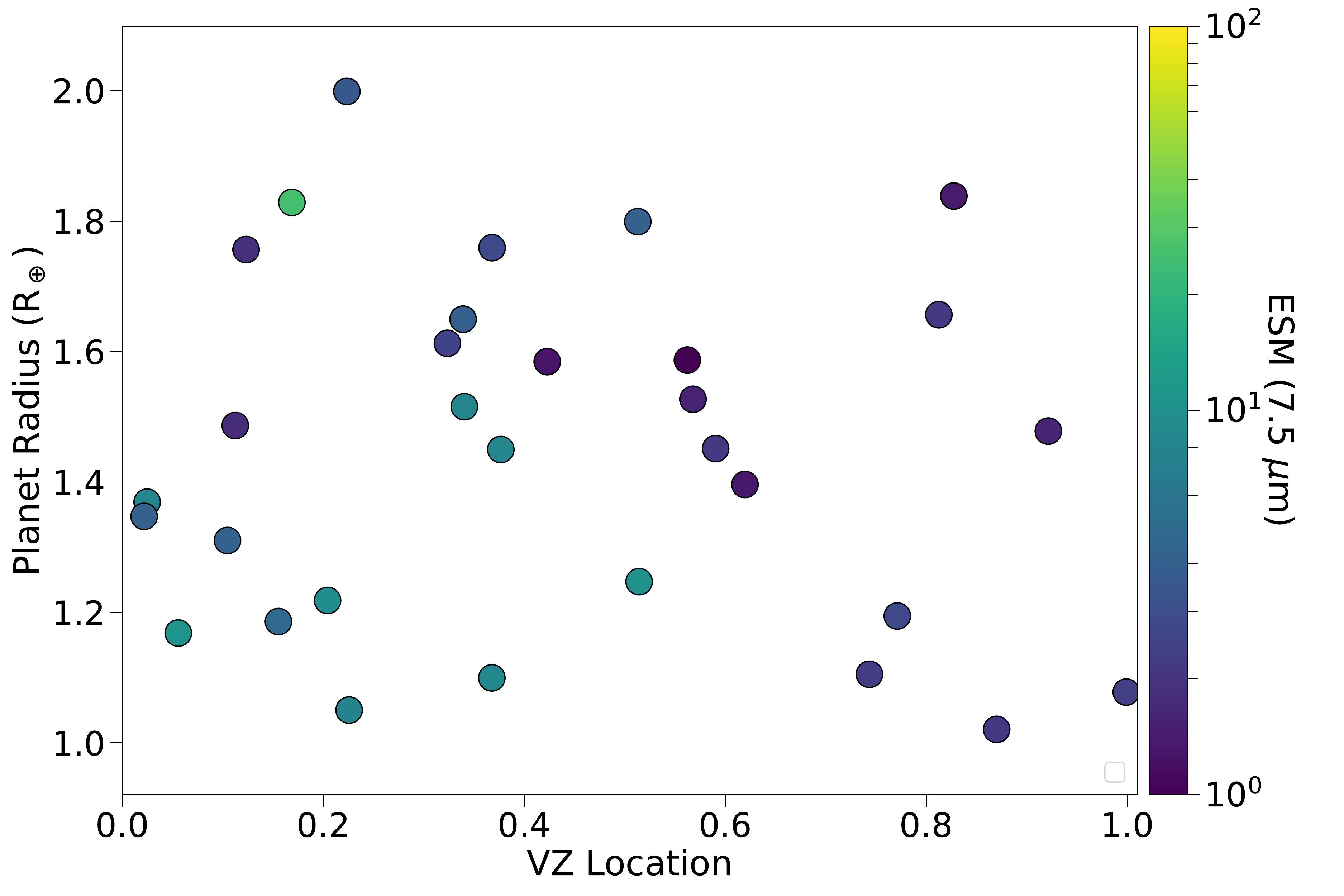}%
}
\subfloat{%
  \includegraphics[width=0.95\columnwidth]{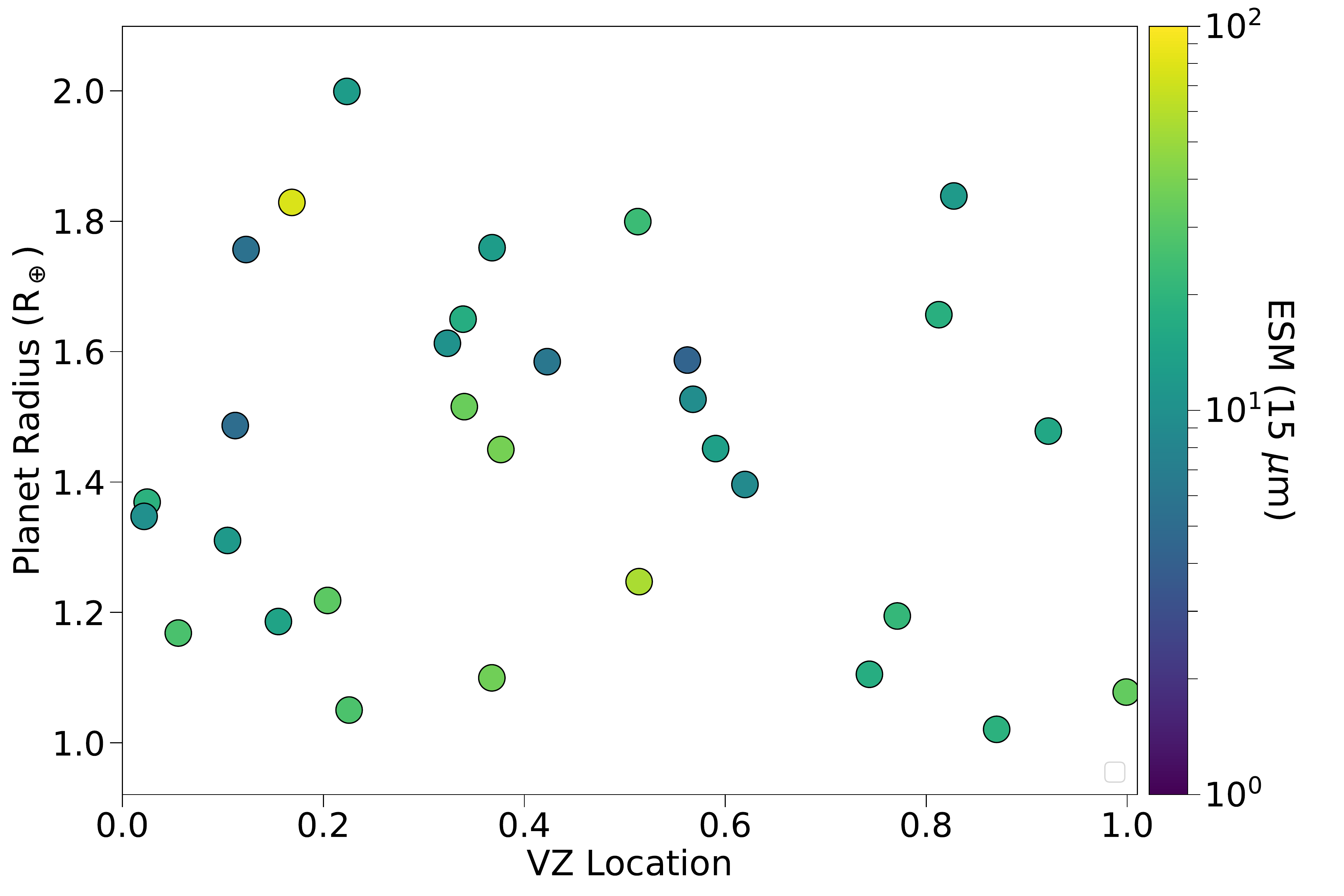}%
}
\caption{The ESM values of all non-transiting, terrestrial VZ planets at both 7.5 $\mu$m (left) and 15 $\mu$m (right). 
\label{fig:ESM}}
\end{figure*}

Limiting terrestrial VZ targets to those that are known to transit excludes 37 planets that may be useful for testing the VZ boundaries and studying evolutionary pathways of VZ planets. The atmospheres of these planets will be inaccessible to JWST, but a concept mission named the Mid Infrared Exoplanet CLimate Explorer \citep[MIRECLE;][]{Mandell2022} would use thermal emission to constrain the surface temperature and atmospheric composition of non-transiting rocky exoplanets. Although MIRECLE is currently a concept, it is valuable to identify which non-transiting VZ planets would be beneficial for studying Venus and the VZ. Non-transiting planet candidates were chosen based on their vicinity to the outer VZ boundary, similarity in size to Venus, and Emission Spectroscopy Metric \citep[ESM;][]{kempton2018} value at both 7.5 and 15 $\mu$m (Figure \ref{fig:ESM}). Similar to TSM, ESM values provide a first-order S/N estimate of the emission spectrum of a planet. A more detailed explanation of both the TSM and ESM can be found in \citet{kempton2018}. Using these criteria we chose six planets: GJ~3323~b \citep{astudillo2017harps}; GJ~1061~b, GJ~1061~c, and GJ~1061~d \citep{dreizler2020reddots}; Proxima Centauri~b \citep{anglada2016terrestrial}; and Teegarden's Star~b \citep{zechmeister2019carmenes}. 

GJ~3323~b has ESM values of 10.1 and 56.0 at 7.5 and 15 $\mu$m, respectively. These ESM values are the second highest among all non-transiting, terrestrial VZ planets. Its orbit places it at roughly the center of the VZ, and it has a measured mass and estimated radius of 2.02 M$_\oplus$ and 1.25 R$_\oplus$, respectively. The planet is a good test case to study the runaway greenhouse limit for more massive planets. 

Proxima Centauri b has a measured mass of 1.27~$M_\oplus$, an estimated radius of 1.05~$R_\oplus$, and an orbital eccentricity of $\sim$0.35 \citep{bixel2017,kane2017a}. Its eccentric orbit causes it to spend 75\% of its orbit in the HZ and 25\% in the VZ, during which it receives varying incident flux that ranges from 0.36--1.55~$F_\oplus$. Studying the planet is valuable for determining whether its short exposure to higher insolation flux was sufficient enough to force it into a runaway greenhouse. This will also inherently be a test for the climate evolution of planets with eccentric orbits that straddle the runaway greenhouse boundary. Its ESM value of 33.3 at 15 $\mu$m makes it a strong candidate for observations with emission spectroscopy. 

Teegarden's Star b is physically similar to Earth as well, with a measured mass and estimated radius of 1.05~$M_\oplus$ and 1.02~$R_\oplus$, respectively. It has an ESM value of 18.8 at 15 $\mu$m, which is second lowest of the 6 non-transiting planets we chose. The planet spends all of its orbit within VZ and is only 0.03~AU from the outer VZ boundary, which makes it an excellent case for testing runaway greenhouse conditions. In addition, Teegarden's Star was determined to have an age greater than 8 Gyr \citep{zechmeister2019carmenes}, allowing for the opportunity to study any correlations between planet age and the onset of the runaway greenhouse effect \citep[][]{foley2018carbon,foley2019habitability,unterborn2022mantle}.

The GJ~1061 system has three planets with eccentric orbits that all spend some portion of their orbits in the VZ. The orbit of GJ~1061~b has an eccentricity e = 0.31, and a semi-major axis in the inner half of the VZ (loc = 0.36). It has the highest equilibrium temperature among the GJ~1061 VZ planets, and therefore has the highest ESM of the 3 planets. GJ~1061~c spends 73\% of its orbit within the VZ and the rest within the conservative HZ. Similarly, GJ~1061~d spends 22\% of its orbit in the VZ and the rest in the conservative HZ. Given their orbits, GJ~1061~b and c both offer an opportunity to study the effect of varying insolation flux on a planet's climate. However it should be noted that the eccentricity values reported for these planets are upper limits \citep{dreizler2020reddots}, and circular orbits would make planet c orbit within only the VZ, and planet d only in the HZ. Despite this, the system as a whole would be a useful sample of planetary atmospheres that span across the VZ of a single star.

\setlength{\belowcaptionskip}{2pt}

%%%%%%%%%%%%%%%%%%%%%%%%%%%%%%%%%%%%%%%%%%%%%%%%%%%%%%%%%%%%%%%%%%%%

\subsection{Direct Imaging of Nearby Stars}
\label{sec:DI}

Their intrinsic brightness and closer orbital distance from their host stars make VZ planets potentially interesting targets for the next generation space-based direct imaging (DI) missions, assuming those VZ planets have similar characteristics to that of Venus. Successful DI of an exoplanet requires the target planet to have planet-to-star flux ratios that are above the instrument's minimum contrast ratio requirement at locations in the planet's orbit where the planet-star angular separation is larger than the instrument's inner working angle (IWA). In our estimate of the DI prospects of VZ planets around nearby stars, we assumed usage of the Habitable Exoplanet Observatory (HabEx) mission concept with an external starshade. The HabEx starshade has a required contrast ratio performance of 1$\times$10$^{-10}$, and optimistically, 4$\times$10$^{-11}$ \citep{gaudi2020} at the IWA of 70 mas, which is the angular radius of the starshade itself as seen from the telescope in the visible channel. Typically, planets that orbit inside the IWA of the instrument are considered undetectable due to the obstruction of the starshade. However, it has been demonstrated that this inner limit can be pushed further inward if taking into account the transmittance profile of the starshade itself \citep{gaudi2020,li2021}. For HabEx, such limit, which we refer to as IWA$_{0.5}$, is at angular separation of 56.4 mas from the center of the starshade where the starshade transmittance is only 0.5. Here, we employed the 56.4 mas of IWA$_{0.5}$ and the required 1$\times$10$^{-10}$ contrast threshold as the hard limit of HabEx visibility of VZ planets.

We first checked detectability of known VZ planets taking reported orbital parameters and distance to each star. None of the known VZ planets can be directly imaged owing to the small on-sky angular separation of those planets. Out of these known VZ planets, Proxima Cen b has the largest angular separation of 50 mas, which puts it just inside the HabEx IWA$_{0.5}$ limit of 56.4 mas. Next, we checked whether additional undiscovered VZ planets could be detected through direct imaging. We expanded our VZ planet DI detectability estimate to nearby stars within 30~pc. We queried the Gaia Data Release 3 (DR3) database \citep{gaia2022} and collected all stars within 30~pc with effective temperatures between 2500~K and 7500~K. In total, 3814 nearby stars were collected.

\begin{figure}[htbp]
  \includegraphics[width = \columnwidth]{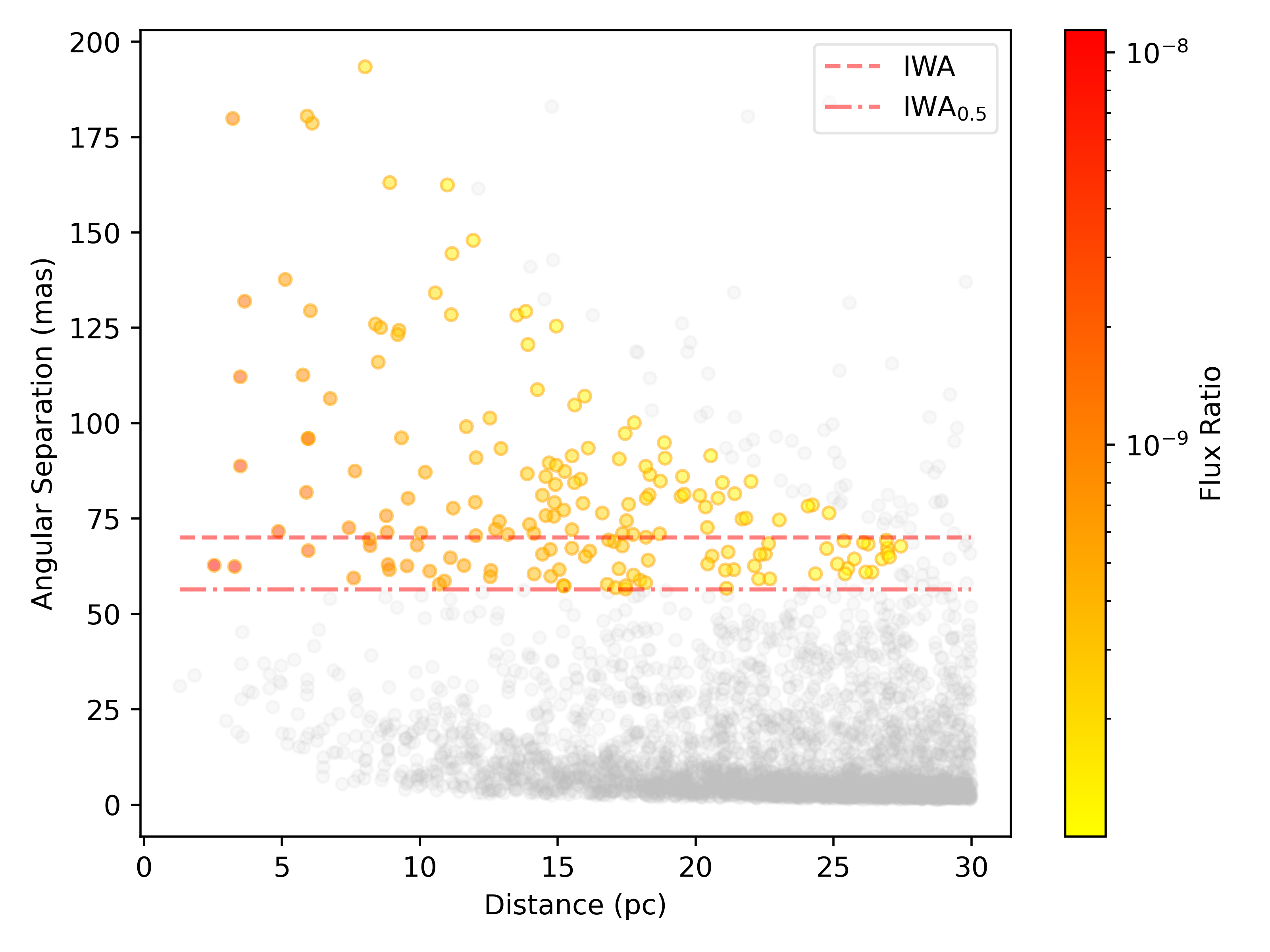}
  \caption{Maximum angular separation of fictitious VZ planets orbiting at the runaway greenhouse boundary around nearby stars with 2500~K $\leq$ $T_\mathrm{eff}$ $\leq$ 7500~K within 30~pc. Face-on inclination was assumed and only planets that orbit outside the IWA/IWA$_{0.5}$ while having flux ratios above the 1$\times$10$^{-10}$ threshold are color coded.}
  \label{fig:VZSize}
\end{figure}

For each star, we calculated the angular separation of the runaway greenhouse boundary \citep{kopparapu2013a,kopparapu2014} based on the distance, stellar radius, and effective temperature values from Gaia DR 3. We injected one fictitious VZ planet with 1 Earth radius at the runaway greenhouse boundary around each star with circular face-on orbit and assumed a geometric albedo similar to that of Venus from the Haystacks modern solar system model, which is 0.6 \citep{roberge2017}. Following \citet{li2021}, we estimated flux ratio variation along the entire orbit for each fictitious planet and determined the maximum planet-to-star flux ratio of the fictitious planet while being outside the IWA$_{0.5}$. In the case of face-on inclination angles, flux ratios of planets are constant since they have the same phase throughout their entire orbit. The planet detectability based on flux ratios were only estimated to the first order. That is, no noise sources such as sky noise, residual starlight from imperfect starshade, clock-induced-charge etc. were included in the estimation. We show the result in Figure~\ref{fig:VZSize}, where the maximum angular separations of fictitious VZ planets are plotted against their distances from Earth to their host stars. Each individual point in the figure represents one fictitious VZ planet orbiting at the runaway greenhouse boundary around one of the nearby stars. Points are color coded based on the planets' maximum flux ratios at any point throughout their entire orbits. VZ planets that are greyed out are therefore undetectable at any orbital locations under our DI setup. Planets that have maximum angular separations smaller than IWA$_{0.5}$ of 56.4 mas or have maximum flux ratios below the instrument minimum contrast threshold of 1$\times$10$^{-10}$ are indicated by grey points. The two red horizontal lines represent the angular sizes of HabEx starshade IWA and IWA$_{0.5}$. 

As can be seen in Figure~\ref{fig:VZSize}, there exhibits a clear downward trend in terms of VZ planets detectability as a function of angular separation. The number of potentially detectable VZ planets around nearby stars quickly diminishes towards the 30 pc distance mark. This is because at larger distances, most VZ planets at the runaway greenhouse boundaries extends smaller than IWA$_{0.5}$ angular separations from their host stars, thus blocked by the starshade. For a few fictitious VZ planets that have large angular separation at large distances (top right corner of Figure~\ref{fig:VZSize}), high stellar effective temperatures push the runaway greenhouse boundaries outward to larger separations and as a result, making planet-star distances larger. Consequently, the planet-to-star flux ratios drop lower because of the large orbital distances. Out of the collected 3814 nearby stars from Gaia, 172 of them may have VZ planets orbiting at the runaway greenhouse boundaries that have large enough angular separations and high enough flux ratios that make them potential future DI targets. Note the face-on inclination Figure~\ref{fig:VZSize} assumed means all fictitious VZ planets on circular orbits would have the same phase and angular separation from their host stars throughout their entire orbits. If we were to increase the assumed orbital inclination towards edge-on, VZ planets whose orbits were outside the IWA$_{0.5}$ from the star would have higher maximum flux ratios at smaller phase angles (i.e., smaller angular separation, nearer to the star and the IWA$_{0.5}$ such that a larger fraction of the planetary disk would be illuminated), which would further increase the number of potential VZ planets detectable by future DI missions. However, note that Figure~\ref{fig:VZSize} is not a complete representation of the DI detectability for all nearby potential VZ planets thanks to the magnitude limitation of Gaia. Because many bright stars from the Gaia catalog suffer from over-saturation, those with Gaia magnitude $G \leq 7$ were not in the Gaia data \citep{gaia2016} and thus were not included in the figure.

%%%%%%%%%%%%%%%%%%%%%%%%%%%%%%%%%%%%%%%%%%%%%%%%%%%%%%%%%%%%%%%%%%%%

\subsection{Contributions to Future Biosignature Searches and Interpretation}
\label{sec:bio}

The search for exoplanet biosignatures is one of the fundamental motivating science goals in the design of future exoplanet observatories \citep{fujii2018,gaudi2020,TheLUVOIRTeam2019}. The most referenced biosignature gas is molecule oxygen (O$_2$), which on Earth is ultimately the result of robust production by photosynthesis organisms \citep{Meadows2018a}. In recent years many possible mechanisms for generating abiotic O$_2$ in planetary atmospheres have been proposed. These abiotic O$_2$ mechanisms include water loss with the selective retention of O$_2$ during a runaway greenhouse \citep[e.g.][]{luger2015b,Schaefer2016,tian2015d} and the decomposition of CO$_2$-rich atmospheres into CO and O$_2$ in the absence of photochemical catalysts such as HO$_x$ species and/or under the influence of photolysis driven by the UV spectral energy distributions of M dwarf hosts \citep[e.g.][]{Gao2015,hu2020o2}. Importantly, these so-called oxygen false positive mechanisms, while supported by theory, remain speculative—we do not yet have observational evidence of their existence. Venus, for example, contains no free O$_2$ despite evidence for water loss from the D/H ratio of atmospheric water vapor \citep{kane2019d}. In the case of Venus, O$_2$ sinks must have outcompeted the net O$_2$ source from H loss, though this is just one data point. Most of these proposed abiotic O$_2$ mechanisms would be most efficiently realized for terrestrial planets that orbit in the VZ, where insolation and presumably water loss is more efficient than for planets in the traditional HZ \citep{luger2015b}. The VZ then provides a novel opportunity to advance the search for exoplanet biosignatures by testing for the existence of some of the most probable false positives on planets that are the least likely to be inhabited. If O$_2$ is not found on VZ planets, it may be a more compelling biosignature on HZ worlds. In contrast, if O$_2$ is common on VZ worlds, we must be more circumspect with biosignature interpretations on planets in the HZ.  Abundant abiotic O$_2$ on VZ planets could be detected via O$_2$-O$_2$ collisionally-induced absorption features at 1.06 and 1.27 $\mu$m in transmission spectroscopy and these and additional visible absorption features in reflected light observations \citep{Lincowski2018,Schwieterman2016,Schwieterman2018}. The best candidates to be observed in search for O$_2$ signatures are those closest to the outer VZ boundary since they are less likely to have experienced hydrodynamic loss of oxygen in their atmospheres.

%%%%%%%%%%%%%%%%%%%%%%%%%%%%%%%%%%%%%%%%%%%%%%%%%%%%%%%%%%%%%%%%%%%%

\section{Conclusions}
\label{sec:conclusions}

In this work we presented a broad overview of the known terrestrial planets in the VZ and provided both measured and estimated data for each. There are currently 317 known terrestrial planets in the VZ, and TESS will continue to provide additional VZ planet candidates as its second extended mission is planned to end in 2025. At the time of writing, the VZ is most heavily populated towards the inner VZ boundary. Although these planets would, in theory, be more amenable to atmospheric observations, their high equilibrium temperature and incident flux make them less useful for comparison to Venus and more susceptible to atmospheric loss. Observations of the atmospheres of VZ planets near the outer VZ boundary present an opportunity to simultaneously study planets which may help understand Venus' past, and test the location of the outer VZ and inner HZ boundaries. 

We listed the eight terrestrial VZ planets which are planned to be observed by JWST, including two pairs of neighboring planets from the TRAPPIST-1 and TOI-776 systems. Observations of multiple VZ planets in the same system can give insight into how differences in planet size, planet mass, and incident flux can result in differences in climate over time without needing to factor in the effects of different host stars. The other JWST VZ targets have unique properties, but all can provide valuable information for understanding the different evolutionary pathways of VZ planets. 
However, the ability of JWST to obtain information about the atmospheres of VZ planets will be highly dependent on their atmospheric composition and cloud coverage, as a Venus-like atmosphere is likely to prevent the detection of any molecular absorption with the time allotted for each planet. However, even a non-detection of molecules in any of the VZ target atmospheres is still valuable since it will illustrate the sensitivity of the JWST instruments and inform future target selection of VZ planets.

Aside from the JWST VZ targets, we chose an additional six terrestrial VZ planets to be considered for future JWST observations. The planets were chosen based on their TSM value, location in the VZ, and size. LTT 1445 A c and L 98-59 d were chosen since studying observations of their atmospheres will be particularly valuable for comparison to their neighboring VZ planets, which are already planned to be observed by JWST. In general, observations of the five suggested planets can be used to investigate how the onset of a runaway greenhouse, or location of the outer VZ boundary, may be dependent on planet mass, incident flux, and orbital eccentricity. In response to the development of the mission concept, MIRECLE, that can study the atmospheres of non-transiting exoplanets, we also provided six non-transiting VZ planets that cannot be probed by JWST. In comparison to the five transiting targets, almost all of the non-transiting targets are more similar to Venus in size and are located closer to the outer VZ boundary. Their locations make them prime targets for testing the outer VZ boundary, and the previously mentioned mission concept would provide the means to make these planets accessible.

The study of Venus-like exoplanets has become increasingly important given the announcement of several future in-situ Venus missions like DAVINCI+ \citep{garvin2022revealing}, VERITAS \citep{cascioli2021determination}, Venera-D \citep{zasova2019venera, vorontsov2011prospective}, and EnVision \citep{widemann2020envision}. The data gathered from these missions will provide invaluable information about Venus, such as updates on the structure and composition of its atmosphere, determining its water loss history, and compiling a high resolution map of its surface. These missions will inherently benefit exoplanetary science since their data will be used to improve climate models capable of simulating Venus-like surface conditions, which will enhance our ability to predict the climates of VZ planets. Modeling the climates of an assortment of VZ planets will strengthen our understanding of what forced Venus into a runaway greenhouse, test the location of the VZ boundaries, and potentially justify the possibility of a temperate period in Venus' past. In particular if many planets in the VZ are found to have Earth-like surface conditions, it will illustrate that the differences in insolation flux between Earth and Venus is not the primary reason for the divergence between the two planets. All of these results would benefit the study of planetary habitability and help identify future targets for the search for life in the universe. 

%%%%%%%%%%%%%%%%%%%%%%%%%%%%%%%%%%%%%%%%%%%%%%%%%%%%%%%%%%%%%%%%%%%%

%%%%%%%%%%%%%%%%%%%%%%%%%%%%%%%%%%%%%%%%%%%%%%%%%%%%%%%%%%%%%%%%%%%%

\floattable
\startlongtable
\begin{deluxetable*}{l c c c c c c c c c c c}
  \tablewidth{0pc}
 \tabletypesize{\scriptsize}
  \centering
  \tablecaption{Venus Zone Planet Properties.
  \label{tab:planets}}
\tablehead{\\[-2ex]
\colhead{Planet}  & \colhead {CHZ (\%)} & \colhead{VZ (\%)} &  \colhead {Loc} & \colhead{$R_p$ \tablenotemark{*} ($R_\oplus$)} & \colhead{$M_p$ \tablenotemark{*} ($M_\oplus$)}  &  \colhead{$P$ (days)} & \colhead{$e$ \tablenotemark{**}} & \colhead{$a$ (AU)} & \colhead{$T_\mathrm{eq}$ (K)} & \colhead{TSM} & \colhead{$K$ (m/s)} }
\startdata
Kepler-367 c       & 0   & 100 & 0.437  & 1.20          & \textit{1.75} & 53.58  & \textit{0} & 0.253 & 375 & 0.83   & -     \\
Kepler-243 c       & 0   & 100 & 0.008  & 1.99          & \textit{4.84} & 20.03  & \textit{0} & 0.142 & 613 & 3.38   & -     \\
Kepler-261 c       & 0   & 100 & 0.068  & 1.99          & \textit{4.94} & 24.57  & \textit{0} & 0.156 & 553 & 6.73   & -     \\
Kepler-327 d       & 0   & 100 & 0.270  & 1.72          & \textit{4.10} & 13.97  & \textit{0} & 0.090 & 427 & 7.75   & -     \\
Kepler-1651 b      & 0   & 100 & 0.180  & 1.84          & \textit{4.45} & 9.88   & 0.13       & 0.072 & 471 & 24.50  & 2.070 \\
Kepler-37 c        & 0   & 100 & 0.008  & 0.75          & \textit{0.35} & 21.30  & \textit{0} & 0.140 & 613 & 5.00   & 0.095 \\
Proxima Cen b      & 75  & 25  & -1     & \textit{1.08} & 1.27          & 11.19  & 0.35       & 0.049 & 251 & 190.71 & 1.591 \\
TRAPPIST-1 c       & 0   & 100 & 0.557  & 1.10          & 1.31          & 2.42   & \textit{0} & 0.016 & 340 & 24.27  & 3.106 \\
\enddata
\tablenotetext{*}{Values that are italicized in the planet radius (R$_p$) and mass (M$_p$) columns indicate that they were determined using a mass-radius relationship rather than being from observations.}
\tablenotetext{**}{Planets that did not have a measured eccentricity are denoted by an italicized 0.}
\tablenotetext{}{An extended version of the table is available in electronic version, which provides uncertainties for all relevant parameters.}
\end{deluxetable*}

%%%%%%%%%%%%%%%%%%%%%%%%%%%%%%%%%%%%%%%%%%%%%%%%%%%%%%%%%%%%%%%%%%%%

\floattable
\startlongtable
\begin{deluxetable*}{l c c c c c c c c c}
  \tablewidth{0pc}
 \tabletypesize{\scriptsize}
  \centering
  \tablecaption{Venus Zone Host Star Properties.
  \label{tab:stars}}
\tablehead{\\[-2ex]
\colhead{Host Star}  & \colhead {R (R$_\odot$)} & \colhead {M (M$_\odot$)} & \colhead{Teff} & \colhead{logg}  &  \colhead{L (L$_\odot$)} & \colhead{Jmag} &  \colhead {Vmag} & \colhead{[Fe/H] (dex)} & \colhead{Dist}}
\startdata
CD Cet           & 0.17 & 0.16 & 3130 & 4.93 & 0.003 & 8.775  & 13.950 & 0.13  & 8.61    \\
EPIC 201170410   & 0.28 & 0.29 & 3648 & 5.00 & 0.012 & 13.464 & 17.698 & -0.05 & -1.00   \\
EPIC 211822797   & 0.58 & 0.62 & 4057 & 4.70 & 0.082 & 12.277 & 15.133 & 0.20  & 186.20  \\
G 264-012        & 0.30 & 0.30 & 3326 & 4.85 & 0.011 & 8.837  & 13.078 & 0.10  & 16.01   \\
GJ 1061          & 0.16 & 0.12 & 2953 & -    & 0.002 & 7.523  & 12.700 & -0.08 & 3.67    \\
GJ 1132          & 0.21 & 0.18 & 3270 & -    & 0.004 & 9.245  & 13.680 & 0.00  & 12.61   \\
GJ 15 A          & 0.38 & 0.38 & 3607 & 4.87 & 0.022 & 5.252  & 8.090  & -0.34 & 3.56    \\
GJ 251           & 0.36 & 0.36 & 3451 & 4.96 & 0.017 & 6.104  & 9.890  & -0.03 & 5.58    \\
\enddata
\end{deluxetable*}

%%%%%%%%%%%%%%%%%%%%%%%%%%%%%%%%%%%%%%%%%%%%%%%%%%%%%%%%%%%%%%%%%%%%

\section*{Acknowledgements}

The authors acknowledge support from NASA grant 80NSSC21K1797, and KB acknowledges the support of NASA grant 80NSSC20K1529, both of which are funded through the NASA Habitable Worlds Program. T.F. acknowledges support from the University of California President's Postdoctoral Fellowship Program. This research has made use of the NASA Exoplanet Archive, which is operated by the California Institute of Technology, under contract with the National Aeronautics and Space Administration under the Exoplanet Exploration Program. The results reported herein benefited from collaborations and/or information exchange within NASA's Nexus for Exoplanet System Science (NExSS) research coordination network sponsored by NASA's Science Mission Directorate. 

\software{Matplotlib \citep{hunter2007matplotlib},
          NumPy \citep{harris2020array}, 
          SciPy \citep{virtanen2020scipy},
          pandas \citep{reback2020pandas}
          }

%%%%%%%%%%%%%%%%%%%%%%%%%%%%%%%%%%%%%%%%%%%%%%%%%%%%%%%%%%%%%%%%%%%%

\bibliographystyle{aasjournal}
\bibliography{references,nea}

%%%%%%%%%%%%%%%%%%%%%%%%%%%%%%%%%%%%%%%%%%%%%%%%%%%%%%%%%%%%%%%%%%%%

\end{document}